\documentclass[12pt]{amsart}

\usepackage{amsfonts,latexsym,amsmath,amssymb, epsfig,array,
delarray}

\title{The Hamiltonian Structure of Discrete KP Equations}
\author{Ali Ulas Ozgur Kisisel}

\newtheorem{Prop}{Proposition}[section]
\newtheorem{Thm}{Theorem}[section]
\newtheorem{Lem}{Lemma}[section]
\newtheorem{Cor}{Corollary}[section]
\newtheorem{Def}{Definition}[section]

\begin{document}

\numberwithin{equation}{section}
\maketitle

\begin{abstract}{This paper investigates Hamiltonian properties
of the algebro-geometric discretization of KP hierarchy introduced
in \cite{Gie1}. A Poisson bracket is introduced. The system is
related to the periodic band matrix system of \cite{vM-M}. It is
shown that the bracket descends to the latter and endows it with
bi-Hamiltonian structure together with the first bracket already
considered in \cite{vM-M}. On the other hand a bi-Hamiltonian
structure for discrete KP seems to be absent for fundamental
reasons. It is proven that the conserved quantities of both
systems are in involution with respect to the bracket. A
construction relating the bracket to a certain intersection
pairing of cycles on a discrete torus is shown. This pairing is
reminiscent of the intersection pairing in ``string topology''
\cite{C-S}.}
\end{abstract}

\section{Introduction}
\setcounter{footnote}{0}
\renewcommand{\thefootnote}{}
 \footnote{Acknowledgements: This paper is based on the
author's dissertation at U.C.L.A.. I am indebted to my advisor
David Gieseker for his ideas, guidance and support} This paper
includes a study of a certain integrable discretization of the KP
hierarchy. This is an algebro-geometric discretization introduced
by Gieseker in \cite{Gie1}. The system has continuous time and
both space directions discrete, and is periodic in the two space
directions with periods, say, $N$ and $M$ respectively. Thus there
is a set of time dependent functions $A(n,m)(t)$, $B(n,m)(t)$
subject to a hierarchy of nonlinear flows where $(n,m)$ is a point
on the $N$ by $M$ discrete torus. We assume that $N$ and $M$ are
relatively prime.

Given an algebraic curve $\mathcal{X}$ of arbitrary genus $g$ with
certain additional properties and additional data including a line
bundle $\mathcal{L}$ of degree $g$ non degenerate in a suitable
sense, the construction produces corresponding $A(n,m)$,$B(n,m)$.
The discrete KP flows correspond to moving $\mathcal{L}$ in linear
directions on the Jacobian of $\mathcal{X}$, keeping the curve and
the rest of the data fixed. One immediately deduces that the flows
commute and there are many conserved quantities. The construction
is generically invertible, i.e. generic $A(n,m)$,$B(n,m)$ come
from such a curve. This is discussed in \cite{Gr} and \cite{Kis}.

This correspondence generalizes the construction relating
hyperelliptic curves and the periodic Toda lattice which is the
case $M=1$ of discrete KP. It is similar to, and motivated by the
construction of van Moerbeke and Mumford \cite{vM-M}, who show the
correspondence between periodic band matrices and curves of
arbitrary genus with additional data. In fact the
algebro-geometric data for the two systems is almost identical.
The discrete KP system is in some sense a finite cover of the band
matrix system. Considering this lifting has several benefits.

Our main purpose is to describe the Hamiltonian nature of the
discrete KP hierarchy which doesn't manifest itself in the
algebro-geometric picture. In particular we introduce a Poisson
bracket for the system for each value of $N,M$. This generalizes
the so called ``second bracket'' of the periodic Toda hierarchy to
arbitrary $M$. Its definition is non-local in the sense that $A$'s
and $B$'s supported at distant points of the torus often have non
vanishing brackets, contrasting periodic Toda. The definition
involves arithmetic properties of the pair $N,M$; there are
roughly two different cases depending on the mod 2 value of the
number of steps in the Euclidean algorithm of the ordered pair
$(N,M)$.

The Poisson bracket descends to a bracket on the band matrix
system as well. Furthermore, we show that this new bracket and the
``first'' bracket in \cite{vM-M} are compatible. The two brackets
endow the band matrix system with a bi-Hamiltonian structure. One
says that two Poisson brackets are compatible if any linear
combination of them is a Poisson bracket. We may also ask whether
they produce the same set of flows when contracted by the
conserved quantities. If they do, then one says that the system is
bi-Hamiltonian. This useful idea was introduced by Lenard and
Magri. Using this, we prove the commutation of conserved
quantities for the band matrix system under the new bracket. The
conserved quantities are not effected by the lifting process, so
we deduce that the conserved quantities for the discrete KP system
also commute .

A natural object to look for is a first bracket for the discrete
KP system. We prove the nonexistence of such a bracket if we
expect it to have some natural properties. To be precise, we prove
that a first bracket producing polynomial expressions and
descending to the first bracket of the band matrix system doesn't
exist. One hopes that there is an intrinsic reason for this. The
author believes the reason is the following: A linear motion of
the curve doesn't correspond to a linear motion of the variables
in discrete KP, as opposed to the band matrix system. This
suggests that the bi-Hamiltonian property seen in many completely
integrable systems is a specific, linear motion case of a more
general, nonlinear motion of Poisson structures, and the precise
meaning of this to us isn't clear yet. There are other interesting
questions, for instance how the discrete bracket relates to
$\mathcal{W}$ algebras (the bracket for continuous KP is very
closely related to $\mathcal{W}$ algebras. See \cite{Dic}.)

Inspection of the conserved quantities $q_{i}$ reveals a pattern
about the monomials that are the summands of the $q_{i}$. These
bijectively correspond to certain closed cycles or unions of
closed cycles on the discrete torus. They have to obey some
additional conditions which can be explicitly characterized. The
Poisson bracket applied to these monomials translates into an
antisymmetric pairing on these cycles. This suggests that in a
proper context  this should be an intersection pairing. However
the pairing depends on the cycles themselves, not just on their
homology classes. The recent preprint \cite{C-S} on string
topology discusses a strikingly similar pairing, and we think that
this is a discrete analog. The commutation of conserved quantities
gives a theorem on the cycles.

Section 2 discusses the periodic Toda lattice. Everything in this
section is well known, but we think that it is a good introduction
for the general case. It should be remarked that we are writing
the equations in terms of the variables after Flaschka's
transformation, so the equations may not be in their most familiar
form for some readers. Section 3 discusses the discrete KP
hierarchy. Most of the results here are unproven and the proofs
can be found in \cite{Kis}. Some functions on the discrete torus
are constructed. In Section 4 we introduce the Poisson bracket and
verify that it indeed is one.

Section 5 describes the relation between the discrete KP and band
matrix systems. It is proven here that the bracket descends. Part
of this proof is shifted to appendix 1 since it is too long and
causes a distraction otherwise. Section 6 discusses the structure
of the conserved quantities of the systems, as well as the
bi-Hamiltonian nature of the band matrix system. It is proven that
the conserved quantities are in involution with respect to the
bracket.

Finally, section 7 discusses the combinatorial construction and
the intersection pairing. Some examples for this section are given
in appendix 2.

\section{Prelude: The periodic Toda Hierarchy}
The periodic Toda lattice is a completely integrable system of
differential-difference equations on $2N$ variables
$A(1),...,A(N)$ and $B(1),...,B(N)$. The complete integrability of
this system is implied by the fact that there exists a Poisson
structure $\{\phantom{x},\phantom{x}\}$ on $R^{2N}$ of generic
rank $N-1$, together with $N+1$ almost everywhere independent
polynomials $q_{1}...,q_{N},q_{2N}$ of $A(i),B(i)$ so that for all
$k,l\in \{1,2,...,N,2N\}$:
\begin{equation}\label{eq:qcommute}
\{q_{k},q_{l}\}=0
\end{equation}
and the flow is given by:
\begin{equation}
\begin{split}
\dot{A}(n)=\{A(n),q_{2}\}\\ \dot{B}(n)=\{B(n),q_{2}\} \\
\end{split}
\end{equation}
These relations imply that the $q_{i}$ are conserved quantities of
the flow.

We will prove the assertions above. We start from the defining
equations of the system:
\begin{equation}\label{eq:Todaeqn}
\begin{split}
\dot{A}(n)=B(n)-B(n+1)\\
\dot{B}(n)=(A(n)-A(n-1))B(n) \\
\end{split}
\end{equation}
In these equations, the indices are assumed to be in
$\mathbb{Z}/N\mathbb{Z}$, and this encodes the periodicity (e.g.
$A(N+1)=A(1)$ etc.).

The Poisson bracket mentioned above is:
\begin{equation}
\begin{split}
\{A(n),B(n)\}_{1}=-B(n)\\
\{A(n-1),B(n)\}_{1}=B(n)\\
\end{split}
\end{equation}
It should be understood that the bracket of two coordinate
functions besides the ones above is zero unless the contrary is a
direct consequence of the antisymmetry property of the Poisson
bracket. The bracket is extended by bilinearity and Leibniz rule
to all $C^{\infty}$ functions of $A(i),B(i)$. This bracket
satisfies the Jacobi identity on coordinate functions, therefore
on all functions.

To prove the existence of conserved quantities $q_{i}$, we show
that it is possible to write the equations in Lax form. Let

 \begin{equation}\label{eq:todal}
L=\begin{bmatrix} -A(1) & 1     & 0   & & 0 &-B(1)/\alpha\\
                -B(2)   & -A(2) & 1   & 0   &    & 0  \\
                  0     & -B(3) & -A(3)& 1  & 0  &     \\
                      &       &   &   &     &    \\
                   ... &       & ...&  & ... &    \\
                      &       &    &  &     &    \\
                \alpha&       &    & 0  &-B(N)&-A(N)
\end{bmatrix}
\end{equation}

and
\begin{equation}
B=\begin{bmatrix} 0 &     &   &   & 0 & -B(1)/\alpha \\
               -B(2)& 0   &   &   &   &  0                   \\
                  0 &-B(3)& 0 &   &   &                    \\
                    &   &   &   &   &                      \\
               ...  &   &...&   &...&                      \\
                    &   &   &   &   &                      \\
                    &   &   & 0 &-B(N)& 0
\end{bmatrix}
\end{equation}
where $\alpha \in \mathbb{C}$ is a free parameter. Then equations
\eqref{eq:Todaeqn} are equivalent to the matrix equation

\begin{equation}
\dot{L}=[B,L]
\end{equation}
and it is a well known result  \cite{Lax} that if $L$ evolves
under an equation of this form, its spectrum is conserved.
Calculating the eigenvalues of $L$ from the equation $\det(L-\beta
I)=0$, one obtains
\begin{equation}
\alpha+(\beta^{N}+q_{1}\beta^{N-1}+...+q_{N})+\frac{q_{2N}}{\alpha}=0
\end{equation}

For any given $\alpha$, the coefficients $q_{i}$ of the polynomial
can be expressed as symmetric polynomials in the roots
$\beta_{j}$. It follows that the $q_{i}$'s must be conserved.
Moreover, for given $A(i),B(i)$, this equation describes a
hyperelliptic plane curve; the coordinate functions of the plane
being $\alpha$ and $\beta$. Paraphrasing the discussion above, we
deduce that this curve is invariant under the flow. It is called
the Bloch spectrum of the periodic Toda system. Considering the
degree of the polynomial in $\beta$, one would expect the genus of
this curve for generic $A(i),B(i)$ to be $N-1$. This is indeed
true. See \cite{Kis} for a proof of a more general statement.

It can be verified that $q_{1}$ and $q_{2N}$ are Casimirs for the
bracket $\{\phantom{x},\phantom{x}\}$ (that is, their Poisson
bracket with any other function is zero). The contraction of the
bracket with $q_{3},...,q_{N}$ give $N-2$ additional flows which
commute with the original flow as well as among themselves in view
of \eqref{eq:qcommute} . The collection of these flows is called
the periodic Toda hierarchy. Let us turn to the algebro-geometric
picture for a moment. The hyperelliptic curve has an associated
Jacobian variety, a complex torus of complex dimension equal to
the genus of the curve (therefore generically $N-1$). The
coordinates of an eigenvector of $L$, if properly normalized, give
$N$ meromorphic functions on the curve. There is a natural line
bundle construction from the divisorial data of these functions
which gives a corresponding point on the Jacobian for generic
$A(i),B(i)$. Under this correspondence, the flows of the Toda
hierarchy precisely correspond to linear flows on the Jacobian,
which commute in virtue of their linearity. In other words, the
algebro-geometric picture provides a linear view of the nonlinear
flows.

There is a second Poisson bracket
$\{\phantom{x},\phantom{x}\}_{2}$ under which the $q_{i}$ are in
involution, and whose contraction with $q_{1},q_{2},...,q_{N-1}$
give back the flows of the Toda hierarchy. $q_{N}$ and $q_{2N}$
are Casimirs for this new bracket. From now on we denote the first
bracket by $\{\phantom{x},\phantom{x}\}_{1}$. The definition of
$\{\phantom{x},\phantom{x}\}_{2}$ is:

\begin{equation}\label{eq:todabrac2}
\begin{split}
\{A(n),A(n+1)\}_{2}=B(n+1)\\
\{A(n),B(n)\}_{2}=A(n)B(n)\\
\{A(n-1),B(n)\}_{2}= -A(n-1)B(n)\\
\{B(n-1),B(n)\}_{2}=-B(n-1)B(n)\\
\end{split}
\end{equation}
$\{\phantom{x},\phantom{x}\}_{1}$ and
$\{\phantom{x},\phantom{x}\}_{2}$ are compatible brackets which
means that any linear combination of the two brackets is a Poisson
bracket. Furthermore,

\begin{equation}\label{eq:biham}
\{q_{i},\cdot\}_{1}=\{q_{i-1},\cdot\}_{2}
\end{equation}
for $i=2,...,N$. The two brackets are said to form a Poisson pair
for the Toda hierarchy, and systems having such pairs are called
bi-Hamiltonian. The relation \eqref{eq:biham} automatically
implies relations \eqref{eq:qcommute} which may otherwise be
difficult to prove.

\section{The Discrete KP hierarchy}
\subsection{Description of the system}

From here on, assume that $N$ and $M$ are positive integers such
that $\gcd(N,M)=1$. The algebro-geometric discretization of the KP
equation that we will discuss was introduced by Gieseker in
\cite{Gie1}. We want to describe the system through construction
of the generalization of $L$ in \eqref{eq:todal}. Consider the
following problem: We look for functions $\Psi(n,m,t)$, where
$(n,m,t) \in \mathbb{Z}\times \mathbb{Z}\times \mathbb{C}$, so
that $\Psi$ is almost periodic in the two space directions of the
lattice $\mathbb{Z}\times \mathbb{Z}$, i.e.:

\begin{equation}
\begin{split}
\Psi(n+N,m)=\alpha\Psi(n,m) \\ \Psi(n,m+M)=\beta\Psi(n,m)
\end{split}
\end{equation}
Moreover, we require that $\Psi(n,m+1)$ (suppressing the time
variable $t$) can be expressed in terms of some of the $\Psi(k,m)$
for all $(n,m)$. More specifically, we require that $\Psi(n,m+1)$
is of the form:

\begin{equation}
\Psi(n,m+1)=\Psi(n+1,m)-A(n,m)\Psi(n,m)-B(n,m)\Psi(n-1,m)
\end{equation}
where $A(n,m)$ and $B(n,m)$ are periodic in both space entries,
with periods $N$ and $M$. Given such a set of $A(n,m), B(n,m)$ the
presence of a nontrivial solution for $\Psi$ forces an algebraic
relation between $\alpha$ and $\beta$. If $M=1$, this reduces to
vanishing of the determinant of the matrix $L-\beta I$ of the
previous section and $\Psi$ becomes an eigenfunction. In the
general case there is a matrix $W$ so that the conditions above
translate as $\Psi \in \ker(W)$. To get $W$, order $\Psi(n,m)$
keeping the second index more significant than the first (i.e. use
the order $(\Psi(1,1),\Psi(2,1),...$ $\Psi(N,1);\Psi(1,2),...)$)
Taking into account the almost periodicity of $\Psi$ as well, one
sees that $W$ is the following $NM$ by $NM$ matrix (presented in
$N$ by $N$ blocks):

\begin{equation} \label{eq:L}
            W = \begin{bmatrix}  -\beta*I_{N}(1) &  0_{N} &   &0_{N}&
X(M)\\
                                     X(1)       & -I_{N}(2) & 0_{N}  &   &
0_{N}\\
                         0_{N}  &    X(2) &   -I_{N}(3)  & 0_{N} &
\\
                            &   &   &   &    \\
                            &...&&...& \\
                            &   &   &   &    \\
                         0_{N}&  &      0_{N}   & X(M-1) & -I_{N}(M)
              \end{bmatrix} \end{equation}
$W$ is in block circulant form. It has two nonzero circulants.
Block $(1,1)$ of $W$ is $-\beta*I_{N}(1)$, and for $i\neq 1$,
block $(i,i)$ is $-I_{N}(i)$. Block $(i+1,i)$ of $W$ is $X(i)$ for
all $i$. (Here, regard $i$ in $\mathbb{Z}/M\mathbb{Z}$.)
$I_{N}(i)$ and $0_{N}$ represent the $N$ by $N$ identity and zero
matrices respectively. The sole purpose of indexing $I_{N}$'s is
making references possible. $X(m)$ is:

\begin{equation}
X(m)=\begin{bmatrix}  -A(1,m)  & 1       &  0  &   & 0 &
-B(1,m)/\alpha
\\
                      -B(2,m)  & -A(2,m) &  1  & 0&    & 0       \\
                         0     & -B(3,m) & -A(3,m)& 1 & 0 &    \\
                               & & & & & \\
                                    &...&   &...&   &...  \\
                               & & & & & \\
                       \alpha  &    0    &     & 0 & -B(N,m) & -A(N,m)
      \end{bmatrix}
\end{equation}
$X(m)$ is a circulant matrix, this time with three nonzero
circulants. The $(i,i)$ entry is $-A(i,m)$, the $(i,i-1)$ entry is
$-B(i,m)$ , and the $(i,i+1)$ entry is $1$ for all $i$ (Here
regard $i$ in $\mathbb{Z}/N\mathbb{Z}$). Notice that this matrix
is the of the form \eqref{eq:todal}.

We label entries of $W$ with two pairs of numbers. The
$((n,m),(k,l))$'th entry where $1\leq n,k \leq N$ and $1\leq m,l
\leq M$  will be the entry $(n,k)$ of block $(m,l)$ of $W$. For
instance $W((n,m),(n,m-1))$ is $-A(n,m-1)$, whereas
$W((n,m),(n-1,m-1))$ is $-B(n,m-1)$.

In order for $W\Psi=0$ and $\Psi$ be nontrivial, $\det(W)$ should
be 0. Given a set of $A(n,m), B(n,m)$, this is the defining
equation of a plane algebraic curve in the variables $\alpha$ and
$\beta$. We saw before that this curve is hyperelliptic for the
periodic Toda system. In the general case, the curve defined by
$\det(W)=0$ has a certain definite behaviour at the $\infty$
points of $\alpha$ or $\beta$. As in the Toda lattice, motion in a
linear direction on the Jacobian of the curve corresponds to
nonlinear evolution equations for $A$ and $B$. These give us the
discrete KP hierarchy. These flows have a large supply of
conserved quantities; the coefficients of $\alpha^{i}\beta^{j}$ in
the curve equation, which in fact are functions of $A(n,m)$ and
$B(n,m)$. We state the correspondence between the
algebro-geometric data and the discrete KP data. The proof of this
correspondence and the unproven results of this section can be
found in \cite{Kis}. Below, $\mathcal{X}$ denotes the
normalization of the curve $\det{W}=0$.

\begin{Thm}\label{thm:correspondence}
There is a natural correspondence between the following sets of data:

1) A generic smooth curve $\mathcal{X}$ of genus $g$ which
possesses points $P,Q$ such that $N(P-Q)=\mathrm{div}(\alpha)$, an
additional list of points $R_{i},S_{i}$, $i=1,...,M$ so that
$M(P+Q)-\sum(R_{i}+S_{i})=\mathrm{div}(\beta)$, where $\alpha$,
$\beta$ are meromorphic functions on $\mathcal{X}$; and a line
bundle $\mathcal{L}$ of degree $g$ on $\mathcal{X}$ such that

\begin{equation}
H^{0}(X,\mathcal{L}((n+m-1)P+(m-n)Q-\sum_{i=1}^{m}(R_{i}+S_{i}))=0
\end{equation}
for all $(n,m)$.

2) Generic functions $A(n,m),B(n,m)$, periodic in the two space
directions with periods $N$ and $M$.

The relation between $g$, $N$ and $M$ is $g=(N-1)M$. The
$\Psi(n,m)$ obtained from $W\Psi=0$, if properly normalized, are
holomorphic sections of the line bundles $\mathcal{L}((n+m)P
+(m-n)Q-\sum_{i=1}^{m}(R_{i}+S_{i}))$.
\end{Thm}

A monomial is said to ``appear'' in the expansion of $\det(W)$ if
there exists a permutation $\pi$ of $NM$ letters so that the
product associated to $\pi$ in the expansion of $\det(W)$ is a
nonzero multiple of this monomial. The ``coefficient'' of a
monomial is the part consisting in $A$'s and $B$'s, as opposed to
the part consisting in $\alpha$ and $\beta$.

\begin{Lem} \label{lem:noncancel}
A monomial appearing in the expansion of $\det(W)$ with a
nonconstant coefficient cannot cancel another monomial with the
same properties.
\end{Lem}

The lemma asserts that the list of $A$,$B$'s in $\mathbf{m}$
determines the associated permutation $\pi$ uniquely, if such a
permutation exists.

\begin{Def}\label{def:degree}
We assign degrees $d$ to multiplicative expressions in
$\alpha,\beta,A,B$ as follows:

\begin{equation}
\begin{split}
 (i)&d(\alpha)=N \\
 (ii)&d(\beta)=M \\
 (iii)&d(A)=1\\
 (iv)&d(B)=2 \\
 (v)&d(c)=0, c\in \mathbb{C}\\
\end{split}
\end{equation}
and the degree of a product is the sum of the degrees.
\end{Def}

The following lemma suggests that this degree assignment is
natural:

\begin{Lem}\label{lem:deg}
If $\mathbf{m}$ is a nonzero monomial appearing in $\det(W)$, then
$d(\mathbf{m})=NM$.
\end{Lem}

There is also a symmetry condition on the monomials that appear:

\begin{Lem}\label{lem:sym}
A monomial of the form  $f(A,B)\alpha^{k}\beta^{j}$ appears in
$\det(W)$ if and only if a monomial of the form
$g(A,B)\alpha^{-k}\beta^{j}$ also does.
\end{Lem}

The following Corollary follows from lemmata \ref{lem:deg} and
\ref{lem:sym}.

\begin{Cor}
A monomial with a coefficient (i.e. $A$,$B$ part) of degree $d$
cannot appear in $\det(W)$ unless $d$ is among the following list
of numbers:

\begin{equation}\label{eq:deg}
\begin{split}
&0\\
&N,N-M,...\\
&2N,2N-M,2N-2M,...\\
&...              \\
&NM,NM-M,NM-2M,NM-3M,...,M,0\\
&...\\
&2NM-2N,2NM-2N-M,2NM-2N-2M,...\\
&2NM-N,2NM-N-M,...\\
&2NM \\
\end{split}
\end{equation}
For $1\leq k\leq M+1$, row $k$ of this list contains the numbers
$(k-1)N-iM$ for $0\leq i \leq \lfloor{\frac{(k-1)N}{M}}\rfloor$
which are all nonnegative. Row $M+1+k$ contains $(M+k)N-iM$ for
$0\leq i \leq \lfloor{\frac{(k-1)N}{M}}\rfloor$, i.e. it has the
same number of entries as row $M+1-k$.
\end{Cor}

It turns out that each of these terms appear in $\det(W)$ for
generic $A,B$. This is easier to prove once we relate the discrete
KP system to the band matrix system. The proof will be given in
section \ref{sec:Ham}.

Using these and some additional information about the monomials,
one can prove theorem \ref{thm:correspondence}. This is discussed
in \cite{Kis}, where also the flow equations have been derived. To
be able to write down the equations, we need to make some
preliminary definitions.

\subsection{Some functions on $ \mathbb{Z}/N\mathbb{Z} \times
\mathbb{Z}/M\mathbb{Z}$} \label{sec:kappa}

As before, suppose that $N,M \in \mathbb{Z}$, and $\gcd(N,M)=1$. Let $S$
denote the set of functions
$f:\mathbb{Z}/N\mathbb{Z} \times \mathbb{Z}/M\mathbb{Z}\rightarrow
\{-1,0,1\}$.

\begin{Prop} \label{prop:kappa}
There is a unique function $\kappa$ in $S$
that satisfies the following conditions:

\begin{equation} \label{eq:kappadiff}
\begin{split}
  (i)&\kappa(0,0)-\kappa(1,-1) =-1 \\
  (ii)&\kappa(0,1)-\kappa(1,0)  =1  \\
  (iii)&\kappa(-1,0)-\kappa(0,-1)=1 \\
  (iv)&\kappa(-1,1)-\kappa(0,0) =-1 \\
\end{split}
\end{equation}
and except for these four values of $(i,j)$, $\kappa(i-1,j+1)=\kappa(i,j)$
\end{Prop}
\noindent \textbf{Proof:} Uniqueness is easy to prove, because if
two such functions exist, their  difference has to be a constant.
But by (i) and (iv) the only possibility for $\kappa(0,0)$ is $0$.
Therefore the constant is zero.

To prove existence, we note
that since $\gcd(N,M)=1$,
$(-1,1)$ is a generator. Look at the sequence

\begin{equation}\label{eq:seq}
(-1,1),(-2,2),...,(-a,a),...
\end{equation}
We will distinguish the two cases below:

1) Suppose in the sequence $\eqref{eq:seq}$, $(1,0)$ appears before
$(-1,0)$. Then we declare

\begin{equation}
\begin{split}
\kappa(-1,1)&=\kappa(-2,2)=...=\kappa(1,0)=-1 \\
\kappa(1,-1)&=\kappa(2,-2)=...=\kappa(-1,0)=1 \\
\end{split}
\end{equation}
and $\kappa(a,b)=0$ if $(a,b)$ is not in these lists.

2) Suppose $(-1,0)$ appears before $(1,0)$. We declare

\begin{equation}
\begin{split}
\kappa(-1,1)&=\kappa(-2,2)=...=\kappa(0,-1)=-1 \\
\kappa(1,-1)&=\kappa(2,-2)=...=\kappa(0,1)=1   \\
\end{split}
\end{equation}
and $\kappa(a,b)=0$ if $(a,b)$ is not in these lists.

One can check that $\kappa$ satisfies the conditions that we asked for.
$\Box$

\noindent \textbf{Remark:} A natural question is which case
happens when. It turns out that the deciding quantity is the
parity of the number of steps in the Euclidean algorithm for the
ordered pair $(N,M)$. In particular we have alternate cases for
$(N,M)$ and $(M,N)$.

Note that $\kappa(n,m)=-\kappa(-n,-m)$.

We will use the following definition only in section
\ref{sec:combinatorial}.

\begin{Def}
A function $f$ in $S$ will be called ``strictly row
alternating'' if: Say $f(n,m)=1$. Let $i$ be the smallest positive integer
such that $f(n+i,m)=1$ again. Then there exists exactly one $0<j<i$ such
that $f(n+j,m)=-1$.
\end{Def}

\begin{Prop} \label{prop:rowalternate}
$\kappa$ is strictly row alternating.
\end{Prop}
\noindent \textbf{Proof:} We will prove this for the second case
in the proof of \ref{prop:kappa}. The other case can be obtained
by transposing everything. First, we remark that if
$\kappa(n,m)=-1$ then $\kappa(n+1,m)=+1$. Indeed, this is true for
$(n,m)=(-1,1)$ by construction. On the other hand, $(-1,1)$ and
$(0,1)$ are the beginning points of a trail of $-1$'s and  a trail
of $+1$'s respectively. There are an equal number of elements in
each trail. So this assertion holds everywhere.  Notice that this
proves the proposition immediately, since if one moves towards the
left starting from a $+1$, the first nonzero number encountered is
a $-1$. A second $-1$ encountered will have a $+1$ as its right
neighbor, which will be encountered before. $\Box$

Next, we would like to define two other functions $\rho,\phi$ in $S$:

\begin{Def}\label{def:rhophi}
\begin{gather}
\rho(n,m)=\kappa(n+1,m)+\kappa(n,m) +
\delta_{(n,m),(0,0)}-\delta_{(n,m),(-1,0)} \\
\phi(n,m)=-\rho(-n-1,-m)-\rho(-n,-m)
\end {gather}
\end{Def}
(Here and later, $\delta$ is the Kronecker delta function, i.e.
$\delta_{X,Y}=1$ if $X=Y$, and $0$ otherwise, etc.)

Then the following holds:

\begin{equation}
\begin{split}
\rho(n-1,m+1)-\rho(n,m)=&(\kappa(n,m+1)-\kappa(n+1,m))\\
&+(\kappa(n-1,m+1)-\kappa(n,m)) \\
&+\delta_{(n-1,m+1),(0,0)}-\delta_{(n-1,m+1),(-1,0)}\\
&-\delta_{(n,m),(0,0)}
+\delta_{(n,m),(-1,0)}\\
\end{split}
\end{equation}
Therefore $\rho$ is the unique function in $S$ satisfying the following
conditions:
\begin{equation}\label{eq:rhodiff}
\begin{split}
  (i)&\rho(-2,0)-\rho(-1,-1) =1 \\
  (ii)&\rho(0,1)-\rho(1,0)  =1  \\
  (iii)&\rho(-1,0)-\rho(0,-1)=-1 \\
  (iv)&\rho(-1,1)-\rho(0,0) =-1  \\
\end{split}
\end{equation}
and $\rho(n-1,m+1)=\rho(n,m)$ for all other $(n,m)$.

\begin{Prop}
$\phi$ is an odd function. Also,
\begin{equation}
\phi(k,l)=\rho(k,l)+\rho(k-1,l)
\end{equation}
\end{Prop}
\noindent \textbf{Proof:}
\begin{equation}
\begin{split}
\phi(k,l)=& -\rho(-k-1,-l)-\rho(-k,-l) \\
         =& -\kappa(-k,-l)-\kappa(-k-1,-l)-\kappa(-k+1,-l)-\kappa(-k,-l)\\
          &
-\delta_{(-k-1,-l),(0,0)}+\delta_{(-k-1,-l),(-1,0)}-\delta_{(-k,-l),(0,0)}+\delta_{(-k,-l),(-1,0)}\\
         =& \kappa(k,l)+\kappa(k+1,l)+\kappa(k-1,l)+\kappa(k,l) \\
          & -\delta_{(k,l),(-1,0)}+\delta_{(k,l),(1,0)} \\
\end{split}
\end{equation}
$\kappa(k+1,l)+\kappa(k-1,l)$ and
$-\delta_{(k,l),(-1,0)}+\delta_{(k,l),(1,0)}$ are both odd, therefore
$\phi$ is odd. The identity can be checked directly.$\Box$

Now we can give the equations for the first flow of the discrete
KP hierarchy. As remarked before, the proof is in \cite{Kis}:

\begin{Prop}
The equations of evolution are
\begin{equation}
\begin{split}\label{eq:evolution}
 \dot{A}(n,m) =& B(n,m)-B(n+1,m)+(\sum\kappa(k-n,l-m)A(k,l))A(n,m) \\
 \dot{B}(n,m) =& (\sum \rho(k-n,l-m)A(k,l))B(n,m) \\
\end{split}
\end{equation}
\end{Prop}

We define further analogs of $\rho$ and $\phi$ to be used in
section \ref{sec:mvm}.

\begin{Def}
Suppose $x \leq y$ are nonnegative integers. Define

\begin{equation}
\zeta^{0,y}(n,m)= \kappa(n+y,m)+\kappa(n+y-1,m)+\dots+\kappa(n,m)
\end{equation}
and
\begin{equation}\label{eq:zetaexpand1}
\begin{split}
\zeta^{x,y}(n,m)=
&\zeta^{0,y}(n,m)+\zeta^{0,y}(n-1,m)+\dots+\zeta^{0,y}(n-x,m) \\
&+\delta_{(n,m),(1,0)}+\delta_{(n,m),(2,0)}+\dots+\delta_{(n,m),(x,0)}\\
&-\delta_{(n,m),(-y,0)}-\delta_{(n,m),(-y+1,0)}-\dots-\delta_{(n,m),(-y+x-1,0)}\\
\end{split}
\end{equation}

If $x>y$, define $\zeta^{x,y}$ by
\begin{equation} \label{eq:zetasym}
\zeta^{x,y}(n,m)=-\zeta^{y,x}(-n,-m)
\end{equation}

\end{Def}
Note that this is a valid definition since $\zeta^{x,x}$ is an odd
function.

One can show that, if $x\leq y$;

\begin{equation}
\begin{split}
\zeta^{x,y}(n,m)=
&\zeta^{x,0}(n,m)+\zeta^{x,0}(n+1,m)+\dots+\zeta^{x,0}(n+y,m) \\
&+\delta_{(n,m),(1,0)}+\delta_{(n,m),(2,0)}+\dots+\delta_{(n,m),(x,0)}\\
&-\delta_{(n,m),(-y,0)}-\delta_{(n,m),(-y+1,0)}-\dots-\delta_{(n,m),(-y+x-1,0)}\\
\end{split}
\end{equation}
and if $x\geq y$
\begin{equation}
\begin{split}
\zeta^{x,y}(n,m)=
&\zeta^{x,0}(n,m)+\zeta^{x,0}(n+1,m)+\dots+\zeta^{x,0}(n+y,m) \\
&+\delta_{(n,m),(x-y+1,0)}+\delta_{(n,m),(2,0)}+\dots+\delta_{(n,m),(x,0)}\\
&-\delta_{(n,m),(-y,0)}-\delta_{(n,m),(-y+1,0)}-\dots-\delta_{(n,m),(-1,0)}\\
\end{split}
\end{equation}
also,
\begin{equation} \label{eq:zetaexpand2}
\begin{split}
\zeta^{x,y}(n,m)=
&\zeta^{0,y}(n,m)+\zeta^{0,y}(n-1,m)+\dots+\zeta^{0,y}(n-x,m) \\
&+\delta_{(n,m),(x-y+1,0)}+\delta_{(n,m),(2,0)}+\dots+\delta_{(n,m),(x,0)}\\
&-\delta_{(n,m),(-y,0)}-\delta_{(n,m),(-y+1,0)}-\dots-\delta_{(n,m),(-1,0)}\\
\end{split}
\end{equation}
So, $\kappa=\zeta^{0,0}$ ,
$\rho=\zeta^{0,1}+\delta_{(n,m),(0,0)}-\delta_{(n,m),(-1,0)}$ ,
$\phi=\zeta^{1,1}$.

We have the following addition rule for $\zeta$

\begin{Prop}
If $x<y$
\begin{equation}
\zeta^{x+1,y}(n,m)=\zeta^{x,y}(n,m)+\zeta^{0,y}(n-x-1,m)+\delta_{(n,m),(x+1,0)}-\delta_{(n,m),(-y+x,0)}
\end{equation}
and if $x\geq y$
\begin{equation} \label{eq:zetatemp3}
\zeta^{x+1,y}(n,m)=\zeta^{x,y}(n,m)+\zeta^{0,y}(n-x-1,m)+\delta_{(n,m),(x+1,0)}-\delta_{(n,m),(-y+x+1,0)}
\end{equation}
\end{Prop}
\noindent \textbf{Proof:} When $x<y$, by $\eqref{eq:zetaexpand1}$,
the left hand side is
\begin{equation} \label{eq:zetatemp1}
\begin{split}
&\zeta^{0,y}(n,m)+\zeta^{0,y}(n-1,m)+\dots+\zeta^{0,y}(n-x-1,m)\\
+&\delta_{(n,m),(1,0)}+\dots+\delta_{(n,m),(x+1,0)}\\
-&\delta_{(n,m),(-y,0)}-\dots-\delta_{(n,m),(-y+x,0)}\\
\end{split}
\end{equation}
whereas the right hand side is
\begin{equation} \label{eq:zetatemp2}
\begin{split}
&\zeta^{0,y}(n,m)+\zeta^{0,y}(n-1,m)+\dots+\zeta^{0,y}(n-x,m)+\zeta^{0,y}(n-x-1,m)\\
+&\delta_{(n,m),(1,0)}+\dots+\delta_{(n,m),(x,0)}\\
-&\delta_{(n,m),(-y,0)}-\dots-\delta_{(n,m),(-y+x-1,0)}\\
\end{split}
\end{equation}
looking at $\eqref{eq:zetatemp1}$ and $\eqref{eq:zetatemp2}$, the
assertion follows

When $x\geq y$, use $\eqref{eq:zetaexpand2}$  to expand terms this time.
Left hand side is
\begin{equation}
\begin{split}
&\zeta^{0,y}(n,m)+\zeta^{0,y}(n-1,m)+\dots+\zeta^{0,y}(n-x-1,m)\\
+&\delta_{(n,m),(x+1-y+1,0)}+\dots+\delta_{(n,m),(x+1,0)}\\
-&\delta_{(n,m),(-y,0)}-\dots-\delta_{(n,m),(-1,0)}\\
\end{split}
\end{equation}
whereas the right hand side is
\begin{equation}
\begin{split}
&\zeta^{0,y}(n,m)+\zeta^{0,y}(n-1,m)+\dots+\zeta^{0,y}(n-x,m)+\zeta^{0,y}(n-x-1,m)\\
+&\delta_{(n,m),(x-y+1,0)}+\dots+\delta_{(n,m),(x,0)}\\
-&\delta_{(n,m),(-y,0)}-\dots-\delta_{(n,m),(-1,0)}\\
\end{split}
\end{equation}
so $\eqref{eq:zetatemp3}$ follows.  $\Box$.

The following formulae, obtained by switching $x$ and $y$ in the
proposition, also hold:

If $y<x$
\begin{equation} \label{eq:zetaadd1}
\zeta^{x,y+1}(n,m)=\zeta^{x,y}(n,m)+\zeta^{x,0}(n+y+1,m)-\delta_{(n,m),(-y-1,0)}+\delta_{(n,m),(-y+x,0)}
\end{equation}
and if $y\geq x$
\begin{equation}
\zeta^{x,y+1}(n,m)=\zeta^{x,y}(n,m)+\zeta^{x,0}(n+y+1,m)-\delta_{(n,m),(-y-1,0)}+\delta_{(n,m),(-y+x-1,0)}
\end{equation}

A unifying feature of all $\zeta$'s is the following property they have:

\begin{Prop} \label{prop:zetaspec}
Say $x\leq y$.
$\zeta^{x,y}$ is the unique function in $S$
satisfying the following conditions:
\begin{equation}
\begin{split}
(i)&\zeta^{x,y}(0,0)-\zeta^{x,y}(1,-1)=-1\\
(ii)&\zeta^{x,y}(x,1)-\zeta^{x,y}(x+1,0)=1\\
(iii)&\zeta^{x,y}(-y-1,0)-\zeta^{x,y}(-y,-1)=1\\
(iv)&\zeta^{x,y}(-y+x-1,1)-\zeta^{x,y}(-y+x,0)=-1\\
\end{split}
\end{equation}
and $\zeta^{x,y}(n+1,m-1)=\zeta^{x,y}(n,m)$ for all other $(n,m)$.

If $x\geq y$, $\zeta^{x,y}$ is the unique function in $S$
satisfying
\begin{equation} \label{eq:zetadiff2}
\begin{split}
(i)&\zeta^{x,y}(-1,1)-\zeta^{x,y}(0,0)=-1\\
(ii)&\zeta^{x,y}(x,1)-\zeta^{x,y}(x+1,0)=1\\
(iii)&\zeta^{x,y}(-y-1,0)-\zeta^{x,y}(-y,-1)=1\\
(iv)&\zeta^{x,y}(-y+x,0)-\zeta^{x,y}(-y+x+1,-1)=-1\\
\end{split}
\end{equation}
and $\zeta^{x,y}(n+1,m-1)=\zeta^{x,y}(n,m)$ for all other $(n,m)$.
\end{Prop}

\section{The Poisson bracket} \label{sec:bracket}

Looking at the form of equations $\eqref{eq:evolution}$ , we guess
a quadratic Poisson bracket for the discrete KP hierarchy. This
section is devoted to introducing this bracket, and to verifying
that it indeed is a Poisson bracket.

\begin{Thm} \label{thm:bracket}
The following bracket $\{\phantom{x},\phantom{x}\} $ is a Poisson
bracket. (We give the formulae on coordinate functions only. It is
extended by bilinearity and Leibniz rule to all
$\mathcal{C}^{\infty}$ functions of $A$'s and $B$'s.)
\begin{equation} \label{eq:bracket}
\begin{split}
 \{A(k,l),A(n,m)\}&=\kappa(k-n,l-m)A(k,l)A(n,m)  \\
  &+\delta_{(k,l),(n-1,m)}B(n,m)-\delta_{(k,l),(n+1,m)}B(n+1,m)\\
 \{A(k,l),B(n,m)\}&=\rho(k-n,l-m)A(k,l)B(n,m) \\
 \{B(k,l),B(n,m)\}&=\phi(k-n,l-m)B(k,l)B(n,m) \\
\end{split}
\end{equation}
Moreover, if $q_{1}=\sum_{k,l}A(k,l)$ then
\begin{gather}
\dot{A}(n,m)=\{q_{1},A(n,m)\}\\
\dot{B}(n,m)=\{q_{1},B(n,m)\}
\end{gather}
\end{Thm}

Before proving the theorem, we prove a preliminary proposition

\begin{Prop}\label{prop:antisym}
Consider a set of functions $X_{i}$, $i\in I$. Suppose
$\{\phantom{x},\phantom{x}\}$ is bilinear and satisfies the
Leibniz rule. Suppose $\{X_{i},X_{j}\}=\mu_{(i,j)}X_{i}X_{j} $
where $\mu_{(i,j)}=-\mu_{(j,i)}$. Then any triple of $X_{i}$'s
satisfies the Jacobi identity.
\end{Prop}
\noindent \textbf{Proof:}
\begin{equation}
\begin{split}
 \{X_{i},\{X_{j},X_{k}\}\} & = \{X_{i},\mu_{(j,k)}X_{j}X_{k}\} \\
 & = (\mu_{(i,j)}\mu_{(j,k)}+\mu_{(i,k)}\mu_{(j,k)})X_{i}X_{j}X_{k} \\
 & = (\mu_{(i,j)}\mu_{(j,k)}-\mu_{(j,k)}\mu_{(k,i)})X_{i}X_{j}X_{k} \\
\end{split}
\end{equation}
So, adding over all cyclic permutations, we get 0. $\Box$

\noindent \textbf{Proof of theorem \ref{thm:bracket}:}  We remark
that bracket $\eqref{eq:bracket}$ is antisymmetric since the
functions $\kappa$ and $\phi$ are odd.

We should verify Jacobi identity for all triplets of $A$'s and $B$'s.
By proposition \ref{prop:antisym} , we need to do this only when two of
the three functions are $A(n-1,m)$ and $A(n,m)$, since this is the only case
that one gets brackets outside the scope of Proposition
\ref{prop:antisym}. By toroidal
symmetry, we don't loose generality assuming $(n,m)=(0,0)$. We shall
consider all possibilities for the third function.

(i)The third function is $A(k,l)$, $(k,l)\neq(-2,0)$ or $(1,0)$;
\begin{equation}
\begin{split}
\{A(k,l),\{A(-1,0),&A(0,0)\}\} +\{A(0,0),\{A(k,l),A(-1,0)\}\}\\
&+\{A(-1,0),\{A(0,0),A(k,l)\}\} \\
=&\{A(k,l),B(0,0)+\kappa(-1,0)A(-1,0)A(0,0)\} \\
&+\{A(0,0),\kappa(k+1,l)A(k,l)A(-1,0)\} \\
&+\{A(-1,0),\kappa(-k,-l)A(k,l)A(0,0)\} \\
=&(\rho(k,l)-\kappa(k+1,l)-\kappa(k,l))A(k,l)B(0,0) \\
&+(\kappa(-1,0)\kappa(k+1,l)+\kappa(-1,0)\kappa(k,l)+\kappa(k+1,l)\kappa(-k,-l)\\
&+\kappa(k+1,l)\kappa(1,0)+\kappa(-k,-l)\kappa(-k-1,-l)+\kappa(-k,-l)\kappa(-1,0))\\
&A(k,l)A(0,0)A(-1,0) \\
\end{split}
\end{equation}
Coefficient of $A(k,l)B(0,0)$ vanishes by definition of $\rho$, and
inspection shows that terms in the other parenthesis cancel in pairs.

(ii) The third function is  $A(1,0)$

\begin{equation}
\begin{split}
\{A(-1,0),\{A(0,0),&A(1,0)\}\} +\{A(1,0),\{A(-1,0),A(0,0)\}\}\\
&+ \{A(0,0),\{A(1,0),A(-1,0)\}\} \\
=&\{A(-1,0),B(1,0)+\kappa(-1,0)A(0,0)A(1,0)\}\\
&+\{A(1,0),B(0,0)+\kappa(-1,0)A(-1,0)A(0,0)\}\\
&+\{A(0,0),\kappa(2,0)A(1,0)A(-1,0)\}\\
=&(\rho(-2,0)-\kappa(-1,0)-\kappa(-2,0))A(-1,0)B(1,0)\\
&+(\rho(1,0)-\kappa(1,0)-\kappa(2,0))A(1,0)B(0,0)\\
&+(\kappa(-1,0)\kappa(-1,0)+\kappa(-1,0)\kappa(-2,0)+\kappa(-1,0)\kappa(2,0)\\
&+\kappa(-1,0)\kappa(1,0)+\kappa(-1,0)\kappa(2,0)+\kappa(2,0)\kappa(1,0))\\
&A(-1,0)A(0,0)A(1,0)
\\
=&0\\
\end{split}
\end{equation}
again, all three parentheses are 0.

We need not consider $A(-2,0)$ since it is analogous to (ii). The
cases $A(0,0)$ and $A(-1,0)$ trivially work.

(iii) The third function is $B(k,l)$
\begin{equation}
\begin{split}
\{B(k,l),\{A(-1,0),&A(0,0)\}\} + \{A(0,0),\{B(k,l),A(-1,0)\}\}\\
&+ \{A(-1,0),\{A(0,0),B(k,l)\}\} \\
=&\{B(k,l),B(0,0)+\kappa(-1,0)A(-1,0)A(0,0)\}\\
&+\{A(0,0),-\rho(-k-1,-l)A(-1,0)B(k,l)\}\\
&+\{A(-1,0),\rho(-k,-l)A(0,0)B(k,l)\}\\
=&(\phi(k,l)+\rho(-k-1,-l)+\rho(-k,-l))B(k,l)B(0,0)\\
&+(-\rho(-k-1,-l)\kappa(-1,0)-\rho(-k,-l)\kappa(-1,0)
-\rho(-k-1,-l)\kappa(1,0)\\
&-\rho(-k-1,-l)\rho(-k,-l)
+\rho(-k,-l)\kappa(-1,0)+\rho(-k,-l)\rho(-k-1,-l))\\
&B(k,l)A(-1,0)A(0,0) \\
=&0\\
\end{split}
\end{equation}
and this finishes the proof.   $\Box$

\noindent \textbf{Remark:} The case $M=1$ gives back the periodic
Toda lattice if $\kappa$, $\rho$, $\phi$ are interpreted in a
degenerate way. These functions were defined via their difference
properties on pairs of points (see $\eqref{eq:kappadiff}$ and
$\eqref{eq:rhodiff}$). Whenever there are two or more conditions
for a pair of points on a function in $S$, impose the sum of them
on the pair. Then $\kappa(n)$ will be $0$ for all $n$,
$\rho(0)=1$, $\rho(-1)=-1$ and $\rho(n)$ is $0$ otherwise,
$\phi(1)=1$, $\phi(-1)=-1$ and $\phi(n)$ is $0$ otherwise. Then
equations $\eqref{eq:evolution}$ become the evolution equations
for the periodic Toda lattice, and $\eqref{eq:bracket}$ reduces to
\eqref{eq:todabrac2}.

\section{Relation with the system of Mumford-Van Moerbeke} \label{sec:mvm}
\subsection{Description of the system}

In their 1979 paper, Mumford and van Moerbeke demonstrate a
correspondence between periodic band matrices and algebraic curves
with additional data \cite{vM-M} (They do not assume the
equivalent of $\gcd(N,M)=1$, or that the curve is smooth, but we
assume these for our discussion. Following notation of
\cite{vM-M}, we assume $M=M^{'}$, and also that the rightmost loop
of the band matrix consists in 1's entirely).

This construction is related to, and was motivational for the
construction of \cite{Gie1}. The algebro-geometric pictures differ
only in one aspect: In \cite{Gie1}, the divisor corresponding to
zeroes of $\beta$ is further broken down into $M$ divisors of
degree 2. (Some dictionary: $\alpha$ here $\equiv$ $h$ in
\cite{vM-M}, $\beta$ here $\equiv$ $z$ in \cite{vM-M}). In other
words, the underlying curves, and functions $\alpha$, $\beta$ are
unaltered. The variables subject to the flows, on the other hand,
differ. It is one of the purposes of this section to show the
relation.

Let us describe the band matrix system via a spectral problem.
Define $L$ to be the linear differential operator:

\begin{equation}
(Ls)(n)=\tilde{c}_{0}(n)s(n+M)+\tilde{c}_{1}(n)s(n+M-1)+...+\tilde{c}_{2M}(n)s(n-M)
\end{equation}
Let $T$ be the translation operator $(Ts)(n)=s(n+N)$. Suppose the
coefficients $\tilde{c}_{i}$ are periodic with period $N$, i.e.
$\tilde{c}_{i}(n)=\tilde{c}_{i}(n+N)$. Then $L$ commutes with $T$.
We look for common eigenfunctions of $L$ and $T$. This translates
as vanishing of a determinant as before.

We now compare the matrices for the two systems.

It turns out that it is more convenient to look at infinite
matrices in the $N$ direction (only) in order to compare the two
systems. We do this for $W$ first: in $W$ of display
$\eqref{eq:L}$ , replace each $N$ by $N$ block by the
corresponding infinite periodic matrix of width $3$. Denote the
infinite counterparts of matrices by adding a $\tilde{}$ to the
notation. $\tilde{I}_{N}(j)$ is an infinite identity matrix, and
$\tilde{X}(m)$ becomes an infinite tridiagonal matrix so that

\begin{equation}
\begin{split}
\tilde{X}(m)(k,l)=-A(k,m&) \quad \mathrm{if}\quad k=l  \\
                 -B(k,m&)\quad \mathrm{if}\quad k=l+1 \\
                 &1 \quad\mathrm{if}\quad k=l-1 \\
                 &0 \qquad \mathrm{otherwise}\\
\end{split}
\end{equation}
Here $A(k,l)$ and $B(k,l)$ are periodic in both slots, and the periods
are $N$ and $M$ respectively.

Turning back to the band matrix problem, let $\tilde{C}$ be the infinite
periodic band matrix of width $(2M+1)$ and period $N$ such that
$\tilde{c}_{i}(k)$ is the element $\tilde{C}(k,k+M-i)$ of
$\tilde{C}$. With this particular choice,  $\tilde{c}_{i}$'s
are on the $i$th diagonal. Here we number the diagonals from right to
left so that the main diagonal is always the $M$'th.
Saying that $\tilde{C}$ is of width $(2M+1)$ amounts to saying that
$\tilde{c}_{i}(k)=0$ for $i<0$ or $i>2M$. We furthermore ask that
$\tilde{c}_{0}(k)=1$ for each $k$. The periodicity condition means
$\tilde{c}_{i}(k+N)=\tilde{c}_{i}(k)$, as we assumed above.

The Bloch spectrum is the set of $(\alpha,\beta)$ such that

\begin{equation}
\begin{split}
(\tilde{C}-\beta)s&=0 \\
s(n+N)&=\alpha s(n)  \\
\end{split}
\end{equation}
In order to get the curve equation in the variables $\alpha,\beta$, one
considers the $N$ by $N$ matrix $C-\beta I$, where $C$ is obtained from
$\tilde{C}$ by taking one period. To take periodicity into account,
multiply the lower triangular piece of the band matrix
sticking out by $\alpha$ and translate by $-N$, and multiply the upper
triangular piece sticking out by $\alpha^{-1}$ and translate by $N$. If
there still remain portions sticking out, repeat these operations (see
\cite{vM-M}).

\subsection{The algebraic relation}

The algebraic relation between the two systems \cite{Gie1} and
\cite{vM-M} at the level of matrices (i.e. the relation between
$\tilde{W}$ and $\tilde{C}$) is the following: Use row reduction
to clear block $(1,M)$ of $\tilde{W}$ using block $(M,M)$, which
is $-\tilde{I}_{N}(M)$. This creates a new nonzero block,
$(1,M-1)$ in $\tilde{W}$. We may further clear this new block,
using block $(M-1,M-1)$ this time, and proceed inductively, each
time clearing the new block formed on block-row $1$, using the
next diagonal block in the up left direction. When this process is
over, blocks $(i,i)$ for $i>2$ are still $\tilde{I}_{N}(i)$, but
block $(1,1)$ is a band matrix of width $(2M+1)$ with exactly the
properties described. Except for an extra $-\beta\tilde{I}_{N}(1)$
it contains, this will be the $\tilde{C}$ that corresponds to
$\tilde{W}$.

If the finite matrices $W$ and $C$ are considered instead, it is easily
seen that an analogous reduction gives the analogous result, i.e. the
lower and upper triangular corners acquire the correct power of $\alpha$.
Since row reduction does not change the determinant, we see
that $C-\beta I$ and $W$ have the same determinant, except for a possible
difference in sign coming from the $-I_{N}'s$. Notice that the
reduced matrix is in block-triangular form, hence its determinant is
the product of the determinants of its diagonal blocks. Thus, the curve
equations for the two systems are identical, as we have remarked before,
the new functions $c_{i}$ being certain polynomials in $A$'s and $B$'s.

We want to show that through this series of reductions one can get almost
any set of $\tilde{c}_{i}$ by a suitable choice of $A,B$. In order to
prove this, we take a closer look at the reduction process. We would like
to keep track of all intermediate steps in the process of reducing
$\tilde{W}$ to $\tilde{C}$. Rename $\tilde{C}$ as $\tilde{W}^{(1)}$. Let
$\tilde{W}^{(M+1-j)}$ denote the new block in block-row $1$ of $\tilde{W}$
obtained at the $j$th intermediate step. For
instance, with this notation, $\tilde{W}^{(M)}=X(M)$. We label the entries
of $\tilde{W}^{(j)}$ the same way as for $\tilde{C}$: The entries are
$\tilde{c}^{(j)}_{i}(k)$, where $\tilde{c}^{(j)}_{M+1-j}$ are on the main
diagonal. So, for instance, $\tilde{c}_{i}(k)$ above is
$\tilde{c}^{(1)}_{i}(k)$.

The explicit formula for the reduction from $\tilde{c}^{(j+1)}$ to
$\tilde{c}^{(j)}$ is

\begin{equation} \label{eq:map}
\begin{split}
\tilde{c}^{(j)}_{i}(k)=\tilde{c}^{(j+1)}_{i}(k)&-A(k-i+M+1-j,j)\tilde{c}^{(j+1)}_{i-1}(k)\\
                        &-B(k-i+M+2-j,j)\tilde{c}^{(j+1)}_{i-2}(k)\\
\end{split}
\end{equation}
Let $\mathcal{A}^{(j)}$ denote the affine space with the ring of
functions
$\mathbb{C}[c^{(j)}_{i}(k),A(k,j-1),B(k,j-1)]$ , where $k=1,...,N$
and $i=1...,2(M+1-j)$, and let $\mathcal{B}^{(j)}$ denote the affine space
with the ring of functions $\mathbb{C}[c^{(j)}_{i}(k)]$. Then
$\eqref{eq:map}$ gives a rational map $\phi_{j}$ from
$\mathcal{A}^{(j+1)}$ to $\mathcal{B}^{(j)}$. Notice that these two
varieties have the same dimension.

\begin{Prop}\label{prop:dominant}
$\phi_{j}$ is a dominant map.
\end{Prop}
\noindent \textbf{Proof:} It is enough to check that the
differential of $\phi_{j}$ is surjective at one point (hence in an
open subvariety) of $\mathcal{A}^{(j+1)}$. We calculate the
differential at the point defined by: for all $k$;
$A(k,j)=B(k,j)=0$, $\tilde{c}_{2M-2j}^{(j+1)}(k)=1$, and
$\tilde{c}_{i}^{(j+1)}=0$ for all other $i$. Easy computation
shows:

\begin{equation}\label{eq:diff1}
d\tilde{c}_{i}^{(j)}(k)=d\tilde{c}_{i}^{(j+1)}(k)
\end{equation}
except for $i=1,2,2M-2j+1,2M-2j+2$. And for these four values of $i$,

\begin{equation}\label{eq:diff2}
\begin{split}
d\tilde{c}_{1}^{(j)}(k)=d\tilde{c}_{1}^{(j+1)}(k)-dA(k+M-j,j) \\
d\tilde{c}_{2}^{(j)}(k)=d\tilde{c}_{2}^{(j+1)}(k)-dB(k+M-j,j) \\
d\tilde{c}_{2M-2j+1}^{(j)}(k)=-dA(k-M+j,j) \\
d\tilde{c}_{2M-2j+2}^{(j)}(k)=-dB(k-M+j,j) \\
\end{split}
\end{equation}
From $\eqref{eq:diff1}$ and $\eqref{eq:diff2}$ it is clear that the
differential is surjective. $\Box$

Since each step of the reduction is a dominant map, the overall reduction
from $A,B$'s to $\tilde{c}$'s is a dominant map. Thus we can obtain almost
any set of $\tilde{c}_{i}(k)$ by choosing suitable $A,B$.

\subsection{The bracket for the second system}

Now we turn to the Poisson bracket.

It is evidently possible to calculate
$\{\tilde{c}_{i}(k),\tilde{c}_{j}(l)\}$ in terms
of $A$'s and $B$'s by keeping track of the row operations
$\eqref{eq:map}$. The important result we will prove is that
these brackets can be expressed back in terms of $\tilde{c}$'s only.

\begin{Thm} \label{thm:mvmbracket1}
The bracket of Theorem \ref{thm:bracket} induces a bracket on the set
of variables $\tilde{c}_{i}(k)$, $i=1,\dots,2M$. The formulae for this
bracket are given as follows  (suppose $i_{1} \geq i_{2}$) :

First define

\begin{equation}
f( \tilde{c}_{i_{1}}(k_{1}),\tilde{c}_{i_{2}}(k_{2}) )
=( \delta_{k_{2}\leq
k_{1}}\delta_{k_{2}-i_{2}\leq k_{1}-i_{1}}-\delta_{k_{2} \geq
k_{1}}\delta_{k_{2}-i_{2}\geq
k_{1}-i_{1}} )\tilde{c}_{k_{1}-k_{2}+i_{2}}(k_{1})
\tilde{c}_{k_{2}-k_{1}+i_{1}}(k_{2})
\end{equation}

Then the bracket is:

\begin{equation}\label{eq:mvmbracket1}
\begin{split}
\{\tilde{c}_{i_{1}}(k_{1}),\tilde{c}_{i_{2}}(k_{2})\}  =&
\zeta^{i_{1}-1,i_{2}-1}(k_{1}-k_{2},0)\tilde{c}_{i_{1}}(k_{1})\tilde{c}_{i_{2}}(k_{2})\\
&
+\sum_{l\in\mathbb{Z}}f(\tilde{c}_{i_{1}}(k_{1}),\tilde{c}_{i_{2}}(k_{2}+lN))\\
\end{split}
\end{equation}
\end{Thm}

We first explain what $f$ does in words. Imagine a rectangle placed
on the band matrix, such that its sides are parallel to the rows and
columns of the matrix, and two of the diagonally opposite vertices of the
rectangle sit on top of the points where $\tilde{c}_{i_{1}}(k_{1})$ and
$\tilde{c}_{i_{2}}(k_{2})$ are. Then
$f(\tilde{c}_{i_{1}}(k_{1}),\tilde{c}_{i_{2}}(k_{2}))$ is the product of
$\tilde{c}$'s under the two remaining vertices of the rectangle, with a
coefficient of $-1,0$ or $1$. This product is necessarily zero if the
rectangle is too large compared to the width of the band matrix. So, for
given $i_{1},k_{1}$;
$f(\tilde{c}_{i_{1}}(k_{1}),\tilde{c}_{i_{2}}(k_{2}))$ is
nonzero for only finitely many pairs $i_{2},k_{2}$. In particular the sum
in $\eqref{eq:mvmbracket1}$ is finite.

The proof of the theorem is by induction. A stronger assertion is true: at
every intermediate step of the reduction, we obtain an induced bracket for
the set of variables at that level (This fails if one tries to consider
several levels at once). We need these intermediate steps for the
induction. So we will state and prove a theorem that is slightly stronger
than \ref{thm:mvmbracket1}.

\begin{Thm} \label{thm:mvmprime1}
The bracket of Theorem \ref{thm:bracket} induces a bracket on the set
of variables $\tilde{c}_{i}^{(j)}(k)$, $i=1,\dots,2M-2j+2$ for any given
$1\leq j\leq M$. It is given by the following formulae (suppose
$i_{1}\geq i_{2}$):

\begin{equation}\label{eq:mvmprime1}
\begin{split}
\{\tilde{c}_{i_{1}}^{(j)}(k_{1}),\tilde{c}_{i_{2}}^{(j)}(k_{2})\}  =&
\zeta^{i_{1}-1,i_{2}-1}(k_{1}-k_{2},0)\tilde{c}_{i_{1}}^{(j)}(k_{1})\tilde{c}_{i_{2}}^{(j)}(k_{2})\\
&
+\sum_{l\in\mathbb{Z}}f(\tilde{c}_{i_{1}}^{(j)}(k_{1}),\tilde{c}_{i_{2}}^{(j)}(k_{2}+lN))\\
\end{split}
\end{equation}
\end{Thm}

To prove \ref{thm:mvmprime1}, we start with two lemmata

\begin{Lem} \label{lem:braclem1}
Suppose $M>j_{0}\geq 1$. If $j > j_{0}$,
\begin{equation}
\{A(k_{1},j_{0}),\tilde{c}^{(j)}_{i}(k_{2})\}=\zeta^{0,i-1}(k_{1}-k_{2},j_{0})
A(k_{1},j_{0})\tilde{c}^{(j)}_{i}(k_{2})
\end{equation}
\end{Lem}
\noindent \textbf{Proof:} We do induction on decreasing $j$. The
statement is clear for $j=M$, because in that case the brackets
are $\{A(k_{1},j_{0}),A(k_{2},M)\}$ and
$\{A(k_{1},j_{0}),B(k_{2},M)\}$, and one only needs to check that
the formulas agree with those in $\eqref{eq:bracket}$. For the
induction step, we expand $\tilde{c}^{(j)}$ using
$\eqref{eq:map}$:

\begin{equation}\label{eq:braclem1}
\begin{split}
\{A(k_{1},j_{0}),\tilde{c}^{(j)}_{i}(k_{2})\}&=\{A(k_{1},j_{0}),
\tilde{c}^{(j+1)}_{i}(k_{2})-A(k_{2}-i+M+1-j,j)\tilde{c}^{(j+1)}_{i-1}(k_{2})\\
& -B(k_{2}-i+M+2-j,j)\tilde{c}^{(j+1)}_{i-2}(k_{2})
\} \\
&=\{A(k_{1},j_{0}),\tilde{c}^{(j+1)}_{i}(k)\}\\
&+\{A(k_{1},j_{0}),-A(k_{2}-i+M+1-j,j)\tilde{c}^{(j+1)}_{i-1}(k_{2})\}\\
&+\{A(k_{1},j_{0}),-B(k_{2}-i+M+2-j,j)\tilde{c}^{(j+1)}_{i-2}(k_{2})\}
\\
\end{split}
\end{equation}
The three brackets on the last line produce weighted product terms only.
Because, if the $\tilde{c}$'s on the right side of
the equation are expanded purely in terms of $A$,$B$, none of the terms on
the second slots of the brackets contain an $A(n,j_{0})$. But according to
$\eqref{eq:bracket}$, the only brackets yielding a term other than a
product are of the form $\{A(n,j_{0}),A(n+1,j_{0})\}$.

We are going to show that all three coefficients in these products are the
same, and that they are all $t=\zeta^{0,i-1}(k_{1}-k_{2},j_{0})$. The
expression for $t$ does not involve $j$, therefore by induction
hypothesis, first of the three brackets in $\eqref{eq:braclem1}$
gives $t$. The coefficient from the second bracket is

\begin{equation}\label{eq:braclem2}
\zeta^{0,i-2}(k_{1}-k_{2},j_{0})+\kappa(k_{1}-k_{2}+i-M-1+j,j_{0}-j)
\end{equation}
Since $j > j_{0} \geq 1$, if the argument of $\kappa$ is shifted by
$(M-j,-M+j)$ in steps of $(1,-1)$, none of the critical points in
$\eqref{eq:kappadiff}$ are
trespassed. Therefore the value of $\kappa$
value does not change and
$\eqref{eq:braclem2}$ becomes

\begin{equation}
\zeta^{0,i-2}(k_{1}-k_{2},j_{0})+\kappa(k_{1}-k_{2}+i-1,j_{0})
\end{equation}
and by definition, this is equal to
$\zeta^{0,i-1}(k_{1}-k_{2},j_{0})$, which is $t$.

From the third bracket we get the coefficient

\begin{equation}
\zeta^{0,i-3}(k_{1}-k_{2},j_{0})+\rho(k_{1}-k_{2}+i-M-2+j,j_{0}-j)
\end{equation}
again, we can shift $\rho$ by $(M-j,-M+j)$ without changing its value, and
get
\begin{equation}
\begin{split}
&\zeta^{0,i-3}(k_{1}-k_{2},j_{0})+\rho(k_{1}-k_{2}+i-2,j_{0})  \\
&=\zeta^{0,i-3}(k_{1}-k_{2},j_{0})+\kappa(k_{1}-k_{2}+i-1,j_{0})+\kappa(k_{1}-k_{2}+i-2,j_{0})
\\
&=t\\
\end{split}
\end{equation}
The equality of the form $\rho=\kappa+\kappa$ is valid since $j_{0}\neq
M$.
$\Box$

\begin{Lem}  \label{lem:braclem2}
If $M > j_{0} \geq 1$ and $j>j_{0}$,
\begin{equation}
\{B(k_{1},j_{0}),\tilde{c}^{(j)}_{i}(k_{2})\}=
\zeta^{1,i-1}(k_{1}-k_{2},j_{0})
B(k_{1},j_{0})\tilde{c}^{(j)}_{i}(k_{2})
\end{equation}
\end{Lem}
\noindent \textbf{Proof:} This follows from the previous lemma,
together with the observation that
$\rho(n,m)=\kappa(n,m)+\kappa(n+1,m)$ if $m \neq 0$ mod $M$.
$\Box$

Now we are ready to prove Theorem \ref{thm:mvmprime1} by induction

\noindent \textbf{Proof of thm \ref{thm:mvmprime1} :} Again, we do
induction on decreasing $j$. Throughout we assume $i_{1}\geq
i_{2}+3$. The remaining cases can be proven in the same manner.

\begin{equation} \label{eq:expand1}
\begin{split}
\{\tilde{c}^{(j)}_{i_{1}}(k_{1}),\tilde{c}^{(j)}_{i_{2}}(k_{2})\}  &=
\{\tilde{c}^{(j+1)}_{i_{1}}(k_{1})-A(k_{1}-i_{1}+M+1-j,j)\tilde{c}^{(j+1)}_{i_{1}-1}(k_{1})\\
&-B(k_{1}-i_{1}+M+2-j,j)\tilde{c}^{(j+1)}_{i_{1}-2}(k_{1}),
\\
&\tilde{c}^{(j+1)}_{i_{2}}(k_{2})-A(k_{2}-i_{2}+M+1-j,j)\tilde{c}^{(j+1)}_{i_{2}-1}(k_{2})\\
&-B(k_{2}-i_{2}+M+2-j,j)\tilde{c}^{(j+1)}_{i_{2}-2}(k_{2})
\}\\
\end{split}
\end{equation}
We can expand this bracket using linearity and Leibniz rule. There
are 25 brackets in this expansion. We know how to evaluate each of
these, using either the induction hypothesis, or the lemmata
\ref{lem:braclem1} and \ref{lem:braclem2}. This confronts us with
a straightforward but admittedly very tedious calculation that
takes several pages. On the other hand it is essential, so we give
the rest of the proof in appendix 1. $\Box$

There is no essential difference if we consider $c$'s rather than
$\tilde{c}$'s. Let $\gamma$ be the algebra homomorphism taking
$\tilde{c}$ to the corresponding $c$, i.e. $\gamma$ removes
$\tilde{} $ 's from the variables, and does not change anything
else. Then,

\begin{Thm} \label{thm:mvmbracket1fin}
The bracket of Theorem \ref{thm:bracket} induces a bracket on the set
of variables $c_{i}(k)$, $i=1,\dots,2M$,$k=1,\dots,N$. This
bracket is given as follows  (suppose $i_{1} \geq i_{2}$) :

\begin{equation}\label{eq:mvmbracket1fin}
\begin{split}
\{c_{i_{1}}(k_{1}),c_{i_{2}}(k_{2})\}  =&
\zeta^{i_{1}-1,i_{2}-1}(k_{1}-k_{2},0)c_{i_{1}}(k_{1})c_{i_{2}}(k_{2})\\
&
+\sum_{l\in\mathbb{Z}}\gamma(f(\tilde{c}_{i_{1}}(k_{1}),\tilde{c}_{i_{2}}(k_{2}+lN)))\\
\end{split}
\end{equation}
\end{Thm}

\section{The Hamiltonian nature of the system}\label{sec:Ham}
\subsection{Conserved Quantities}

The Poisson bracket of \eqref{eq:bracket} endows the discrete KP
system with Hamiltonian structure. We will prove that the
coefficients of the curve equation commute under the bracket. The
curve equations are the same for the discrete KP and band matrix
systems except for a polynomial mapping of the variables which was
shown to be dominant in Proposition \ref{prop:dominant}. Therefore
the problem of commutation of conserved quantities is the same for
the two systems since $A$'s $B$'s or $c$'s enter this problem
through conserved quantities only. On the other hand, in other
issues there are significant differences; a sample case will be
discussed in \ref{sec:pair}.

First we want to determine the Casimirs of bracket $\eqref{eq:bracket}$.

\begin{Thm}
If $\beta=0$, $\det(K)$ is a Casimir of $\eqref{eq:bracket}$ for any
value of $\alpha$.
\end{Thm}
\noindent \textbf{Proof:} If $\beta=0$, the determinant of $L$
splits as a product of block determinants. The following equality
holds:

\begin{equation}
|\det(K)|=|\det(X(1))\det(X(2))...\det(X(M))|
\end{equation}
The determinant of a single $X(m)$ is:

\begin{equation}
\det(X(m))=\alpha+Q_{1}(m)+Q_{2}(m)/\alpha
\end{equation}
for certain polynomials $Q_{1}(m), Q_{2}(m)$. $Q_{2}(m)=B(1,m)...B(N,m)$.
Therefore

\begin{equation}
\{A(1,1),Q_{2}(m)\}=(\sum_{n=1}^{N}\rho(1-n,1-m))A(1,1)Q_{2}(m)
\end{equation}
This is $0$, since sum of $\rho(n,m)$ for fixed $m$ over a period of $n$
is zero.

$Q_{1}(m)$ is the sum of the certain monomials in $A,B$. These are:
$a=A(1,m)...A(N,m)$, any other monomial that can be obtained from
$a$ by replacing $A(k_{i}-1,m)A(k_{i},m)$ with $B(k_{i},m)$ for some
sequence of indices $k_{i}$. In the determinant, all of these monomials
that we mentioned appear with the same sign. Indeed, the replacement
operation changes the signature of the permutation that picks the
monomial, but the sign changes for a second time since $(-A)(-A)$ is
replaced by $(-B)(1)$. Now recall that
$\rho(n,l)=\kappa(n+1,l)+\kappa(n,l)$ unless $l=0$. Therefore,
for $m\neq 1$:

\begin{equation}
\{A(1,1),Q_{1}(m)\}=r(\sum_{n=1}^{N}\kappa(1-n,1-m))A(1,1)Q_{1}(m)
\end{equation}
This, again, is zero. Here $r$ is the number of summands in $Q_{1}$.

If $m=1$, we get some non-product terms as well. An expression of
the form $\{A(1,1),B(2,1)TA(N,1)\}$ gives $-B(1,1)B(2,1)T$ (Here,
$T$ is the remaining part of the monomial). This cancels the
non-product monomial coming from $\{A(1,1),B(1,1)A(2,1)T\}$.

Similarly, $\{A(1,1),A(1,1)TA(N,1)\}$ gives a
$-B(1,1)A(1,1)T$; and furthermore a $\{A(1,1),A(1,1)A(2,1)T^{'}\}$ gives
a $A(1,1)B(2,1)T^{'}$. These cancel the extra product terms in
$\{A(1,1),B(1,1)T\}$ and $\{A(1,1),B(1,2)T^{'}\}$ respectively. To verify
these calculations, recall that
$\rho(n,0)=\kappa(n+1,0)+\kappa(n,0)+\delta_{n,0}-\delta_{n-1,0}$.

Combining all of these observations, $\{A(1,1),\det(X(m))\}=0$ for any
$m$, implying that
$\{A(1,1),\det(K)\}=0$ for any $\alpha$. The proof that
$\{B(1,1),\det(K)\}=0$ is very similar to the first part of the proof for
$A$'s. $\Box$

As a result, there are $2M$ Casimirs of $\eqref{eq:bracket}$ among
the conserved quantities. These are precisely the coefficients
attached to terms of the form $\alpha^{k}\beta^{0}$, in other
words, the coefficients whose degrees belong to the leftmost
column of $\eqref{eq:deg}$. These degrees are $N,2N,\dots,2MN$.

The degree function $d$ is naturally defined on the variables $c$ and
$\tilde{c}$ as well, since these are polynomials in $A$'s and $B$'s. A
glance at $\eqref{eq:map}$ will show that they are homogeneous
polynomials, and $d(c_{i}^{(j)}(k))=d(\tilde{c}_{i}^{(j)}(k))=i$.
As remarked, the curve equations for the two systems are
identical. Therefore, the Casimirs for the induced bracket
$\eqref{eq:mvmbracket1fin}$ are also of the same degrees, and there are
$2M$ of them.

Next, we prove that all degrees in $\eqref{eq:deg}$ are assumed by
some conserved quantity. A non-cancelling lemma similar to Lemma
\ref{lem:noncancel} also holds for $C$, with essentially the same
proof. We only state this result:

\begin{Lem} \label{lem:noncancel2}
A nonzero monomial which contains at least one
$c_{i}$ with $i>0$,  and which appears in the expansion of $\det(C-\beta
I)$ by permutations, cannot cancel another one with the same properties.
\end{Lem}

\begin{Prop} \label{lem:complete}
The conserved quantities have exactly the degrees in $\eqref{eq:deg}$.
There are $(N+1)M$ of them, of which $2M$ are Casimirs for the
bracket described.
\end{Prop}
\noindent \textbf{Proof:} By \ref{lem:noncancel2}, it is enough to
display one monomial for each one of the degrees in
$\eqref{eq:deg}$. For the term with degree $(i+M)N$, $-M\leq i\leq
M$, the monomial $\mathbf{m}_{i}= \alpha^{i}\Pi_{k} c_{M-i}(k)$,
which is the product of all elements on one of the circulants,
does the trick. Now we are going to show that $\mathbf{m}_{i}$ can
be modified in a way to include $\beta^{j}$ without  changing the
$\alpha$ exponent, as long as the degree of the coefficient part
of the monomial remains positive. For ease of presentation suppose
$M<<N$, although it is possible to do this construction in
general, considering $\tilde{C}$ instead of $C$.

Consider $M\times M$ square submatrices $S$ of  $C-\beta I$ so that
the main diagonal of $S_{k}$ consists in
$c_{M-i}(k),c_{M-i}(k+1),...c_{M-i}(k+M-1)$ for some $k$.
Then one of the circulants of $S_{k}$ is composed of
$c_{0}$'s and $c_{M}-\beta$'s only. There are $i$ of these $c_{0}$'s and
$M-i$ of the $c_{M}-\beta$'s. In $\mathbf{m}_{i}$, replace the
product $c_{M-i}(k)c_{M-i}(k+1)...c_{M-i}(k+M-1)$ by the product of these
$c_{0}$'s and $-\beta$ part of the $c_{M}-\beta$'s. Clearly, this new
monomial also appears in the expansion of the determinant, since all we
have done is replace the part of the permutation confined to $S_{k}$ by
some other. Therefore, the $\beta$ exponent can be increased by $M-i$
using just this $S_{k}$. It is possible to choose a smaller $l\times l$
submatrix instead of $S$ as well, keeping the $c_{M-i}$ on the diagonal
again. In this case the $\beta$ exponent can be increased by $l-i$. The
maximal number of disjoint $S_{k}$ that we can choose is
$\lfloor\frac{N}{M}\rfloor$. If $l$ is $N-\lfloor\frac{N}{M}\rfloor M$,
there is one $l \times l$ submatrix disjoint from these, as well. Doing
the replacement operation described for each one of these
submatrices, we can increase the exponent of $\beta$ to a total of

\begin{equation}
t=\lfloor\frac{N}{M}\rfloor (M-i)+ (l-i)
\end{equation}
We cannot get a higher exponent of $\beta$, since

\begin{equation}
\begin{split}
d(\alpha^{i}\beta^{t+1})&=iN+(t+1)M \\
                        &=iN+(\lfloor\frac{N}{M}\rfloor (M-i)+(l-i)+1)M \\
                        &>NM \\
\end{split}
\end{equation}
so the highest power of $\beta$ that $\eqref{eq:deg}$ permits
is gotten. It is easy to check that all intermediate powers of $\beta$ can
be obtained as well, by choosing smaller submatrices whenever necessary.
This finishes the proof that each of the degrees in the list are realized
by some conserved quantity.

We have seen that there are $2M$ Casimirs. Finally, we want to check
that there are $(N+1)M$ numbers in $\eqref{eq:deg}$. Remove the leftmost
column and the middle row of $\eqref{eq:deg}$, which together have
$2M+N-1$ elements. Thus we want to show that the remaining list has
$(N-1)(M-1)$ elements. The two mirror symmetric pieces have $M-1$ rows
each. We show that these pieces can be fit together to give an $(N-1)$ by
$(M-1)$ rectangle. To see that, take two copies of the lower piece
instead. Negate the numbers in the second one. Then the $i$th row of the
first copy and $(M-1-i)$th row of the second copy together consist in the
following numbers in arithmetic progression:

\begin{equation}
(M-i)N-M,(M-i)N-2M,...,-iN+M
\end{equation}
Each one of these sequences contains $N-1$ numbers, and since there are
$M-1$ sequences, the claim is established.  $\Box$

For the algebraic independence of these quantities, we refer the reader to
\cite{vM-M}.

From now on, we label the conserved quantities with respect to their
degree. The quantity of degree $d$ will be denoted by $q_{d}$. For
instance, for the $3 \times 2$ system, the conserved quantities are
$q_{1},q_{2},q_{3}, q_{4},q_{6},q_{7},q_{9}$ and $q_{12}$, of which $q_{3},q_{6},
q_{9}$ and $q_{12}$ are Casimirs.

The complete phase space where the flows take place is $2NM$
dimensional. There are $2M$ independent Casimirs,
$q_{N},q_{2N},...,q_{2MN}$. A level set of these $2M$ quantities,
$q_{N}=x_{1},...,q_{2MN}=x_{2M}$ would be $2M(N-1)$ dimensional. There are
$g=(N-1)M$ independent Hamiltonian flows, on the other hand, so this
accounts exactly for a $2(N-1)M$ dimensional symplectic space. Since the
genus is equal to the number of non Casimir conserved quantities, this
shows that our list is complete.

We close this section by proving a relation between $q_{i}$ and $q_{i+M}$.

\begin{Prop}\label{prop:qlink}
For any $i$ such that $q_{i+M}\neq 0$,
\begin{equation}
|q_{i}|=|\sum_{k}\frac{\partial{q_{i+M}}}{\partial{c_{M}(k)}}|
\end{equation}
\end{Prop}
\noindent \textbf{Proof:} If $q_{i+M}$ is the coefficient of the
term $\alpha^{k}\beta^{j}$, then $q_{i}$ is the coefficient of the
term $\alpha^{k}\beta^{j+1}$. Notice that the $c_{M}(k)$ and
$\beta$ occur in the matrix in the form $c_{M}(k)-\beta$ only.
Number the $\beta$'s from $1$ to $N$ just for the sake of the
following sentence: For any monomial containing $\beta(k)$ there
is a corresponding monomial obtained by swapping $\beta(k)$ with
$c_{M}(k)$, and vice versa. The formula directly follows from this
observation. $\Box$

\subsection{The Poisson pair} \label{sec:pair}

There is a natural definition of the degree $d$ of a Poisson bracket with
respect to its action on a pair of monomials $X$,$Y$:

\begin{equation}
d(\{\phantom{x},\phantom{x}\}_{X,Y})=d(\{X,Y\})-d(X)-d(Y)
\end{equation}
In this formula, $d$ is evaluated on monomials as in definition
\ref{def:degree}.

We shall call a Poisson bracket a ``homogenous bracket'' if its
degree with respect to any pair of monomials is the same.
Bracket $\eqref{eq:bracket}$, and consequently, the induced bracket
$\eqref{eq:mvmbracket1}$ are homogenous brackets of degree 0.

Now, in \cite{vM-M} a bracket of degree $-M$ is given. This bracket is the
generalization of the first bracket
for the classical periodic Toda. Our brackets generalize the second
bracket in periodic Toda. In this section, we show that these two
brackets are compatible for the system of \cite{vM-M}.

Since we will have more than one bracket in question from now on, we
denote bracket $\eqref{eq:bracket}$ or $\eqref{eq:mvmbracket1}$ by
$\{\phantom{x},\phantom{x}\}_{2}$.

Suppose $F$,$G$ are polynomials in $\tilde{c}$. Citing \cite{vM-M}, the
first bracket can be written in the following closed form:

\begin{equation}\label{eq:firstbrac}
\{F,G\}_{1}=\mathrm{Tr}(([(\frac{\partial F }{\partial \tilde{C}})^{+},
(\frac{\partial G }{\partial \tilde{C} })^{+}]-
[(\frac{\partial F }{\partial \tilde{C} })^{-},
(\frac{\partial G }{\partial \tilde{C} })^{-}])\tilde{C}^{T})
\end{equation}
We define the terms in this expression.
Here, $\frac{\partial F}{\partial \tilde{C}}$ denotes differentiation with
respect to the matrix entries of $\tilde{C}$, where the result is a
periodic band matrix with entries as the partial derivatives. This
operation just gives the elementary matrix with  1's in the place of
$\tilde{c}_{i}(k)$ if $F$ is the coordinate
function $\tilde{c}_{i}(k)$. For any
matrix $R$, $R^{+}$ and $R^{-}$ mean:

\begin{equation}
\begin{split}
(R^{+})_{i,j}&=R_{i,j} \quad \mathrm{if} \quad i<j \\
(R^{+})_{i,j}&=0  \qquad \mathrm{if} \quad i\geq j \\
\end{split}
\end{equation}
and $R^{-}=R-R^{+}$

\noindent \textbf{Remark:} Literally speaking, this is slightly
different from the bracket in \cite{vM-M}. The difference is
caused by a change of basis that is explained in \cite{vM-M} page
120. It corresponds to conjugating $\tilde{C}$ by a periodic
diagonal matrix.

Looking at equation $\eqref{eq:firstbrac}$, we see that
the only cases that $\{\tilde{c}_{i}(k),\tilde{c}_{j}(l)\}_{1}$
may be nonzero are those satisfying the two conditions below:

(i) Both of $\tilde{c}_{i}(k)$ and $\tilde{c}_{j}(l)$ are strictly upper
triangular entries of $\tilde{C}$, or both of them are (not necessarily
strictly) lower triangular entries of $\tilde{C}$.

(ii) There is a rectangle with two opposite vertices sitting on
$\tilde{c}_{i}(k)$ and $\tilde{c}_{j}(l+sN)$ for some $s\in \mathbb{Z}$,
and one of the remaining two vertices on a diagonal entry
$\tilde{c}_{M}(h)$. (We allow degenerate rectangles, where one sidelength
is zero)

If (i) and (ii) happen to be true, then the bracket
$\{\tilde{c}_{i}(k),\tilde{c}_{j}(l)\}_{1}$ is the sum of all entries
under the fourth vertices of rectangles that fit the description in
(ii), with their proper signs.

This description resembles the second part of the equation
$\eqref{eq:mvmbracket1}$. This is not accidental.

\begin{Thm} \label{thm:bracrel}
The two brackets, $\{\phantom{x},\phantom{x}\}_{1}$ and
$\{\phantom{x},\phantom{x}\}_{2}$ are compatible. Moreover, on the
generators $c_{i}$,

\begin{equation}
\{\phantom{x},\phantom{x}\}_{1}=\{\phantom{x},\phantom{x}\}_{2}|_{c_{M}(i)}-\{\phantom{x},\phantom{x}\}_{2}|_{c_{M}(i)+1}
\end{equation}
\end{Thm}
\noindent \textbf{Proof:} If we want to compute
$\{c_{r}(i),c_{s}(j)\}_{1}$ where neither $r$ nor $s$ is $M$, we
notice that changing $c_{M}(i)$ to $c_{M}(i)+1$ does not have any
effect on the product term for $\{\phantom{x},\phantom{x}\}_{2}$.
Therefore, for such $r$,$s$, the statement can be verified by
merely looking at the non-product part of
$\{\phantom{x},\phantom{x}\}_{2}$, and observing that the
difference matches $\{\phantom{x},\phantom{x}\}_{1}$. When $r$ or
$s$ is $M$, the product term of $\{\phantom{x},\phantom{x}\}_{2}$
may potentially effect things. The following lemma shows that the
correct thing happens.

\begin{Lem}
If $x\geq M$,
\begin{equation}
\zeta^{x-1,M-1}(n,0)=0
\end{equation}
and if $x<M$
\begin{equation}
\zeta^{x-1,M-1}(n,0)=\delta_{n,-M+1+x}-\delta_{n,0}
\end{equation}
\end{Lem}
\noindent \textbf{Proof:} We begin by showing

\begin{equation}
\zeta^{0,M-1}(n,0)=\delta_{n,-M+1}-\delta_{n,0}
\end{equation}
We claim that $\zeta^{0,M-1}(n,m)$ is $+1$ for $M$ values of $(n,m)$ and
$-1$ for $M$ values of $(n,m)$. Consider the function $g$ of $S$ such that
$g(n,m)=1$ for

\begin{equation}
(n,m)=(-M+1,0),(-M+2,-1),\dots,(0,1)
\end{equation}
and $g(n,m)=-1$ for

\begin{equation}
(n,m)=(-M+1,-1),(-M+2,-2),\dots,(0,0)
\end{equation}
Notice that these sequences are of length $M$, and therefore they cross
row $m=0$ at only one point each.

Now $g$ satisfies the conditions in proposition \ref{prop:zetaspec} for
$x=0$, $y=M-1$, so $g=\zeta^{0,M-1}$.

Thus if we define
\begin{equation}
\tilde{\zeta}^{0,i}(n,m)=\zeta^{0,i}(n,m)-\delta_{(n,m),(-i,0)}+\delta_{(n,m),(0,0)}
\end{equation}
then $\tilde{\zeta}^{0,M-1}(n,0)=0$ for all $n$.

But from $\eqref{eq:zetaexpand2}$ we get that, for $x\geq M$

\begin{equation}
\begin{split}
\zeta^{x-1,M-1}(n,0)&=
\tilde{\zeta}^{0,M-1}(n,0)+
\tilde{\zeta}^{0,M-1}(n-1,0)+\dots+
\tilde{\zeta}^{0,M-1}(n-x+1,0) \\
&=0 \\
\end{split}
\end{equation}
and from $\eqref{eq:zetaexpand1}$, for $x<M$

\begin{equation}\label{eq:almost}
\begin{split}
\zeta^{x-1,M-1}(n,0)=&
\tilde{\zeta}^{0,M-1}(n,0)+
\tilde{\zeta}^{0,M-1}(n-1,0)+\dots+
\tilde{\zeta}^{0,M-1}(n-x+1,0)\\
&+\delta_{n,-M+1+x}-\delta_{n,0} \\
=&\delta_{n,-M+1+x}-\delta_{n,0}\\
\end{split}
\end{equation}  $\Box$

This lemma shows that no product terms arise in
$\{c_{M}(k),c_{i}(j)\}_{2}$ for $i\geq M$. The two $\delta$
terms in $\eqref{eq:almost}$
cancel  the additional non-product terms that come from
$\{c_{M},c_{j}\}_{2}$'s, namely $c_{M}(k)c_{j}(k)$,
$c_{M}(k)c_{j}(k-M+j)$ etc., for $j<M$. One checks that these agree with
$\{\phantom{x},\phantom{x}\}_{1}$ as well. $\Box$

\begin{Thm}
The band matrix system is bi-Hamiltonian. The Hamiltonians for
each row of $\eqref{eq:deg}$ are linked among themselves. The
equations are:
\begin{equation}
\{q_{i+M},\cdot\}_{1}=\{q_{i},\cdot\}_{2}
\end{equation}
The Casimirs for bracket $\{\phantom{x},\phantom{x}\}_{1}$ are the
conserved quantities with degrees on the rightmost of each row in
$\eqref{eq:deg}$.
\end{Thm}
\noindent \textbf{Proof:} The formula is a consequence of
proposition \ref{prop:qlink} and theorem \ref{thm:bracrel}. If
$q_{i}$ is a quantity with degree a rightmost element of
$\eqref{eq:deg}$, then
$\frac{\partial{q_{i}}}{\partial{c_{M}(k)}}=0$ for any $k$.
Therefore we may formally set $q_{i-M}=0$, and
$\{q_{i},\cdot\}_{1}=\{q_{i-M},\cdot\}_{2}=0$. $\Box$

Now we can show that the conserved quantities of both systems are in
involution with respect to $\{\phantom{x},\phantom{x}\}_{2}$.

\begin{Thm} \label{thm:commute}
The conserved quantities
commute with respect to $\{\phantom{x},\phantom{x}\}_{2}$.
\end{Thm}
\noindent \textbf{Proof:} As we have remarked several times
before, proving this statement for the band matrix system proves
it for the discrete KP system as well. We take advantage of the
fact that the former is bi-Hamiltonian. The following reasoning is
standard for bi-Hamiltonian systems:

\begin{equation}
\begin{split}
\{q_{i},q_{j}\}_{2}&=\{q_{i},q_{j+M}\}_{1} \\
                   &=\{q_{i-M},q_{j+M}\}_{2} \\
                   &=... \\
                   &=\{q_{i-kM},q_{j+kM}\}_{2} \\
\end{split}
\end{equation}
The right hand side vanishes eventually, for instance when $k$ is large
enough so that $i<kM$. $\Box$

We finish this section by fulfilling a promise made earlier about
a sample situation that reveals the difference between two
systems. A natural question to ask is: ``What is the first bracket
for the original system?''. Our answer is, there doesn't exist
one, at least one that would be expected naturally. Here is the
exact statement of what we can prove:

\begin{Thm}
Unless $M=1$ or $2$, there does not exist a polynomial Poisson bracket on
$A$,$B$ that descends to the first bracket
$\{\phantom{x},\phantom{x}\}_{1}$ on the
$\tilde{c}_{i}$.
\end{Thm}
\noindent \textbf{Proof:} Assume to the contrary that there exists
one. Any polynomial bracket can be graded with respect to degree.
The highest degree portion of a bracket is again a Poisson
bracket. Indeed, the Jacobi identity for this portion does not
involve lower degree terms, and the Leibniz rule is not effected
by grading in any case anyway. We claim that the highest degree
portion of the candidate bracket $\{\phantom{x},\phantom{x}\}^{'}$
is of degree $-M$. Indeed, since it does not vanish identically on
$A$,$B$, it cannot vanish on $\tilde{C}$ identically either, by
\ref{prop:dominant}. Therefore its degree has to match the degree
of $\{\phantom{x},\phantom{x}\}_{1}$, which is $-M$. On the other
hand, $A$ is of degree 1 and $B$ is of degree $2$. Therefore,if
$M\geq 5$, any degree $-M$ bracket on $A$,$B$ vanishes. The
remaining cases are $M=1,2,3,4$. If $M=3$ or $4$, $\{A,A\}=0$ by
degree, and $\{B,A\}$,$\{B,B\}$ is at most linear in $A$'s, and
certainly cannot contain $B$'s. But the flow equation
$\eqref{eq:evolution}$ does not vanish even if all $A=0$,
therefore $\{\phantom{x},\phantom{x}\}$ cannot give this flow. So
these cases are ruled out. $\Box$

\section{Toroidal Pipe Diagrams} \label{sec:combinatorial}

In this section, we present a combinatorial view of the Poisson
bracket applied to certain functions of $A$,$B$'s and draw some
consequences from \ref{thm:commute}. This interpretation, besides
having some visual appeal, we believe, may be pointing towards a
more fundamental construction in discrete geometry. We set
$B(n,m)=0$ for all $(n,m)$. The Poisson bracket, when applied to
monomials which are summands in the conserved quantities $q_{i}$,
give an intersection pairing of certain ``cycles'' on the discrete
torus $\mathbb{Z}/N\mathbb{Z}\times\mathbb{Z}/M\mathbb{Z}$. One
needs to consider the actual cycles, it is not invariant on the
homology classes of the cycles. This pairing is reminiscent of the
intersection pairing in the context of ``string topology'',
studied in the works of Goldman, Turaev, Chas and Sullivan.

Throughout this section we set $B(n,m)=0$ for all $(n,m)$.
Corresponding to each monomial which is a summand in a conserved
quantity $q_{i}$, we construct a discrete cycle satisfying certain
properties, which will be called a ``toroidal pipe diagram''. We
prove that the correspondence is bijective. Then the theorem about
the commutation of conserved quantities translates into a theorem
about these objects.

\begin{Def}
A ``toroidal pipe diagram'' on $\mathbb{Z}/N\mathbb{Z}\times
\mathbb{Z}/M\mathbb{Z}$ is an assignment of one of three types of
local pictures, or a blank picture to each point of this discrete
torus that obeys the following rules:

$\bullet$ Each point of $\mathbb{Z}/N\mathbb{Z}\times
\mathbb{Z}/M\mathbb{Z}$ is assigned a ``left-and-down knee'', or
an ``up-and-right knee'', or a ``horizontal piece'', or nothing .
A point may be assigned one left-and-down knee and one
up-and-right knee simultaneously, but no other combinations of
multiple assignments to one point are allowed.

$\bullet$ The diagram obtained by joining the abutting ends of the pieces
is closed.
\end{Def}

We abbreviate toroidal pipe diagrams as $\mathcal{TPD}$'s.

\begin{Def}
The degree of a toroidal pipe diagram is the number of horizontal
pieces that it contains.
\end{Def}

\begin{Prop}
Set $B(n,m)=0$. There is a one to one correspondence between the
summands in the conserved quantity $q_{i}$ and toroidal pipe
diagrams of degree $i$. The map is as follows: a horizontal piece
at $(n,m)$ corresponds to a factor $A(n,m)$ in the summand. The
places of the knees are uniquely determined by the horizontal
pieces.
\end{Prop}
\noindent \textbf{Proof:} Set $B(k,l)=0$ in $W$. We want to show
that a product of $A$'s is the coefficient of some
$\alpha^{i}\beta^{j}$ in the expansion of the determinant of $W$
iff the corresponding horizontal pieces are precisely those of a
$\mathcal{TPD}$. We present an algorithm to draw the
$\mathcal{TPD}$ corresponding to a given product in the expansion.
Suppose $A(n,m)$ is in the coefficient. Place a horizontal piece
at $(n,m)$. There are two possibilities: $A(n+1,m)$ is either in
the coefficient or not. If it is, place another horizontal piece
at $(n+1,m)$ and continue from here inductively. Otherwise, place
a left-down knee at $(n+1,m)$. Now, there is only one element that
can be picked in column $(n+1,m)$ since the other two are ruled
out. This is the diagonal element $((n+1,m),(n+1,m))$. Place an
up-right knee at $(n+1,m-1)$. There will be no horizontal piece
assignment to this point later, because the element from row
$(n+1,m)$ is picked, and it is not $A(n+1,m-1)$. Next look at
column $(n+2,m-1)$. There are two possibilities for this column,
$A(n+2,m-1)$ or the diagonal element $((n+2,m-1),(n+2,m-1))$. If
$A(n+2,m-1)$ is picked, we place a horizontal piece at
$(n+2,m-1)$, and we are back at the beginning situation. If it is
not picked, we place a left-down knee at $(n+2,m-1)$ and continue
as before. This shows that we can always continue to the right
without violating the conditions of a $\mathcal{TPD}$. Any
connected component of the diagram has to close up because there
are finitely many points.

Reversing the algorithm, one gets a monomial in the determinant
corresponding to a given $\mathcal{TPD}$. $\Box$

By this correspondence, the Poisson bracket of two
$\mathcal{TPD}$'s is naturally defined. Since we have set all
$B(k,l)$ to zero, all such brackets give products with certain
coefficients. We claim that there exists a straightforward method
to compute this coefficient from the $\mathcal{TPD}$ picture, by
looking at how certain pieces of the two diagrams overlap:

\begin{Prop}
\begin{equation} \label{eq:bractpd}
\{\mathcal{TPD}_{1},\mathcal{TPD}_{2}\}=k\mathcal{TPD}_{1}\mathcal{TPD}_{2}
\end{equation}
where $k$ is given by
\begin{equation}\label{eq:k}
\begin{split}
k=
\#\{(n,m)\|&\mathcal{TPD}_{1} \quad\mathrm{has\; a\; horizontal\; piece\;
 at}\quad
(n,m), \\
           & \mathcal{TPD}_{2}\quad \mathrm{has\; a\; left-and-down\; knee\;
 at}\quad
(n,m)\}\\
-\#\{(n,m)\|&\mathcal{TPD}_{1}\quad \mathrm{has\; a\; horizontal\; piece\;
 at}\quad (n,m),\\
&\mathcal{TPD}_{2} \quad \mathrm{has\; an\; up-and-right\; knee\;
at}\quad (n,m)\}\\
\end{split}
\end{equation}
\end{Prop}
\noindent \textbf{Proof:} The only thing that needs to be verified
is that $k$ is given by equation $\eqref{eq:k}$, since we already
know that the bracket gives the product of the two monomials with
a coefficient. From $\eqref{eq:bracket}$, $k=\sum
\kappa(n-i,m-j)$, where $(n,m)$, $(i,j)$ run over the loci of
horizontal pieces of $\mathcal{TPD}_{1}$ and $\mathcal{TPD}_{2}$
respectively. Fix $(n,m)$. We claim that $\sum\kappa(n-i,m-j)$ is
$0$ unless $\mathcal{TPD}_{2}$ has a knee at $(n,m)$. This follows
from the strict row-alternation property of $\kappa$, as was shown
in proposition \ref{prop:rowalternate}: Remember that
$\kappa(r,s)\neq\kappa(r+1,s-1)$ only for the four values of
$(r,s)$ in $\eqref{eq:kappadiff}$. So if $(n,m)$ is such that none
the places where $(n-i,m-j)$ changes rows correspond to these four
values, $\sum\kappa(n-i,m-j)$ becomes the sum of an alternating
sequence of $1$'s and $-1$'s, which closes up, therefore it should
be zero. It is easy to check that actually two of these four
values contribute to the sum, and the contribution happens
precisely when $(n,m)$ is a knee of $\mathcal{TPD}_{2}$. $\Box$

In $\eqref{eq:bractpd}$, we isolate the coefficient $k$ and define
it to be the ``intersection number'' of the two $\mathcal{TPD}$'s.
We denote this by

\begin{equation}
<\mathcal{TPD}_{1},\mathcal{TPD}_{2}>=k
\end{equation}
The following Proposition and Theorem are only restatements of
results proven above; first follows from the Poisson bracket being
antisymmetric, and the second from the commutation of conserved
quantities. Define the product of two $\mathcal{TPD}$'s as the
union of the two diagrams. Such a product is not necessarily a
$\mathcal{TPD}$ anymore.

\begin{Prop}
The pairing $<\phantom{x},\phantom{x}>$ is antisymmetric.
\end{Prop}

\begin{Thm}
\begin{equation}
\sum_{
\deg(\mathcal{TPD}_{1})=d_{1},
\deg(\mathcal{TPD}_{2})=d_{2},
\mathcal{TPD}_{1}\times\mathcal{TPD}_{2}=\mathrm{fixed}}
<\mathcal{TPD}_{1},\mathcal{TPD}_{2}>=0
\end{equation}
\end{Thm}

\begin{Cor}\label{cor:tpd}
If $<\mathcal{TPD}_{1},\mathcal{TPD}_{2}>\neq 0$, then there
exists at least one other pair of toroidal pipe diagrams
$\mathcal{TPD}_{3},\mathcal{TPD}_{4}$ such that
\begin{equation}
\begin{split}
&\mathcal{TPD}_{1}\times\mathcal{TPD}_{2}=\mathcal{TPD}_{3}\times\mathcal{TPD}_{4}\\
&\deg(\mathcal{TPD}_{1})=\deg(\mathcal{TPD}_{3}) \\
&\deg(\mathcal{TPD}_{2})=\deg(\mathcal{TPD}_{4}) \\
\end{split}
\end{equation}
\end{Cor}

%\subsection{Examples}

%\end{spacing}

\clearpage

\appendix
\noindent \textbf{APPENDIX 1:Completion of the Proof of Theorem
\ref{thm:mvmprime1}}

First, we calculate the coefficients of all weighted product terms
that arise in \eqref{eq:expand1}. Namely, we forget the $f$ terms
in the bracket formulae, and look at what happens to the rest. We
expand $\eqref{eq:expand1}$ by bilinearity into $9$ brackets. In
all derivations (i)-(ix) below, coefficient calculations will be
handled regarding the first indices in $\mathbb{Z}/N\mathbb{Z}$.
For instance $\delta_{a,b}$ is $1$ precisely when $a\equiv b$ mod
$N$, i.e. when $a=b+lN$ for some integer $l$. This saves us from
extra complication in the  notation. We will translate back to
$\mathbb{Z}$ after (ix). Below, we calculate the coefficients
only; the product part of the actual bracket is the product of the
coefficient and the two monomials in question.

(i) Coefficient of product term from
$\{\tilde{c}^{(j+1)}_{i_{1}}(k_{1}),\tilde{c}^{(j+1)}_{i_{2}}(k_{2})\}$
is $\zeta^{i_{1}-1,i_{2}-1}(k_{1}-k_{2},0)$ by induction
hypothesis.

(ii)
$\{\tilde{c}^{(j+1)}_{i_{1}}(k_{1}),-A(k_{2}-i_{2}+M+1-j,j)\tilde{c}^{(j+1)}_{i_{2}-1}(k_{2})\}$
gives
\begin{equation}
\begin{split}
 &\zeta^{i_{1}-1,i_{2}-2}(k_{1}-k_{2},0)
-\zeta^{0,i_{1}-1}(k_{2}-i_{2}+M+1-j-k_{1},j) \\ =
&\zeta^{i_{1}-1,i_{2}-2}(k_{1}-k_{2},0)
+\zeta^{i_{1}-1,0}(k_{1}+i_{2}-M-1+j-k_{2},-j) \\ =
&\zeta^{i_{1}-1,i_{2}-2}(k_{1}-k_{2},0)
+\zeta^{i_{1}-1,0}(k_{1}+i_{2}-M-1+j-k_{2},M-j) \\ =
&\zeta^{i_{1}-1,i_{2}-2}(k_{1}-k_{2},0)
+\zeta^{i_{1}-1,0}(k_{1}+i_{2}-1-1-k_{2},1) \\ =
&\zeta^{i_{1}-1,i_{2}-2}(k_{1}-k_{2},0)
+\zeta^{i_{1}-1,0}(k_{1}+i_{2}-1-k_{2},0)
+\delta_{k_{1}+i_{2}-2-k_{2},i_{1}-1}-\delta_{k_{1}+i_{2}-2-k_{2},-1}
\\
=&\zeta^{i_{1}-1,i_{2}-1}(k_{1}-k_{2},0)+\delta_{k_{1}-k_{2},-i_{2}+1}
-\delta_{k_{1}-k_{2},-i_{2}+2+i_{1}-1} \\
&+\delta_{k_{1}+i_{2}-2-k_{2},i_{1}-1}-\delta_{k_{1}+i_{2}-2-k_{2},-1}
\\ = &\zeta^{i_{1}-1,i_{2}-1}(k_{1}-k_{2},0) \\
\end{split}
\end{equation}
Here, line $1$ to line $2$ is because of $\eqref{eq:zetasym}$.
Line $2$ to line $3$ is by the periodicity of $\zeta$ with period
$M$ in the second variable. Line $3$ to line $5$ is by property
$\eqref{eq:zetadiff2}$ of $\zeta$. Line $5$ to line $6$ is by the
addition rule $\eqref{eq:zetaadd1}$.

In all of the remaining calculations, these and the other results
of section \ref{sec:kappa} will be used repeatedly and freely. For
brevity, sometimes several of them are used at one step.

(iii)$\{\tilde{c}^{(j+1)}_{i_{1}}(k_{1}),-B(k_{2}-i_{2}+M+2-j,j)\tilde{c}^{(j+1)}_{i_{2}-2}(k_{2})\}$
gives
\begin{equation}
\begin{split}
&\zeta^{i_{1}-1,i_{2}-3}(k_{1}-k_{2},0)
-\zeta^{1,i_{1}-1}(k_{2}-i_{2}+M+2-j-k_{1},j) \\
=&\zeta^{i_{1}-1,i_{2}-3}(k_{1}-k_{2},0)
+\zeta^{i_{1}-1,1}(k_{1}-k_{2}+i_{2}-2-M+j,M-j) \\
=&\zeta^{i_{1}-1,i_{2}-3}(k_{1}-k_{2},0)
+\zeta^{i_{1}-1,1}(k_{1}-k_{2}+i_{2}-2,0)
+\delta_{k_{1}-k_{2}+i_{2}-2,i_{1}}-\delta_{k_{1}-k_{2}+i_{2}-2,0}
\\ =&\zeta^{i_{1}-1,i_{2}-3}(k_{1}-k_{2},0)
+\zeta^{i_{1}-1,0}(k_{1}-k_{2}+i_{2}-2,0)
+\zeta^{i_{1}-1,0}(k_{1}-k_{2}+i_{2}-1,0) \\
&+\delta_{k_{1}-k_{2}+i_{2}-2,i_{1}-1}
-\delta_{k_{1}-k_{2}+i_{2}-2,-1}
+\delta_{k_{1}-k_{2}+i_{2}-2,i_{1}}-\delta_{k_{1}-k_{2}+i_{2}-2,0}
\\ =&\zeta^{i_{1}-1,i_{2}-2}(k_{1}-k_{2},0)
+\zeta^{i_{1}-1,0}(k_{1}-k_{2}+i_{2}-1,0)
+\delta_{k_{1}-k_{2},-i_{2}+2}  \\
&-\delta_{k_{1}-k_{2},-i_{2}+3+i_{1}-1}
+\delta_{k_{1}-k_{2}+i_{2}-2,i_{1}-1}
-\delta_{k_{1}-k_{2}+i_{2}-2,-1}  \\
&+\delta_{k_{1}-k_{2}+i_{2}-2,i_{1}}-\delta_{k_{1}-k_{2}+i_{2}-2,0}
\\ =&\zeta^{i_{1}-1,i_{2}-1}(k_{1}-k_{2},0)
+\delta_{k_{1}-k_{2},-i_{2}+1}
-\delta_{k_{1}-k_{2},-i_{2}+2+i_{1}-1} \\
&+\delta_{k_{1}-k_{2},-i_{2}+2}
-\delta_{k_{1}-k_{2},-i_{2}+3+i_{1}-1}
+\delta_{k_{1}-k_{2}+i_{2}-2,i_{1}-1}
-\delta_{k_{1}-k_{2}+i_{2}-2,-1} \\
&+\delta_{k_{1}-k_{2}+i_{2}-2,i_{1}}-\delta_{k_{1}-k_{2}+i_{2}-2,0}
\\ =&\zeta^{i_{1}-1,i_{2}-1}(k_{1}-k_{2},0)
\end{split}
\end{equation}

(iv)$\{-A(k_{1}-i_{1}+M+1-j,j)\tilde{c}^{(j+1)}_{i_{1}-1}(k_{1}),\tilde{c}^{(j+1)}_{i_{2}}(k_{2})\}$
gives
\begin{equation}
\begin{split}
&\zeta^{0,i_{2}-1}(k_{1}-i_{1}+M+1-j-k_{2},j)+
\zeta^{i_{1}-2,i_{2}-1}(k_{1}-k_{2},0) \\
=&\zeta^{0,i_{2}-1}(k_{1}-i_{1}+M+1-j-k_{2},-M+j)+
\zeta^{i_{1}-2,i_{2}-1}(k_{1}-k_{2},0) \\
=&\zeta^{0,i_{2}-1}(k_{1}-i_{1}+1-k_{2},0)+
\zeta^{i_{1}-2,i_{2}-1}(k_{1}-k_{2},0)
+\delta_{k_{1}-i_{1}+1-k_{2},0}-\delta_{k_{1}-i_{1}+1-k_{2},-i_{2}}
\\ =&\zeta^{i_{1}-1,i_{2}-1}(k_{1}-k_{2},0)
+\delta_{k_{1}-k_{2},-i_{2}+1+i_{1}-1} \\
&-\delta_{k_{1}-k_{2},i_{1}-1} +\delta_{k_{1}-i_{1}+1-k_{2},0}
-\delta_{k_{1}-i_{1}+1-k_{2},-i_{2}} \\
=&\zeta^{i_{1}-1,i_{2}-1}(k_{1}-k_{2},0)
+\delta_{k_{1}-i_{1},k_{2}-i_{2}}
-\delta_{k_{1}-i_{1}+1,k_{2}-i_{2}} \\
\end{split}
\end{equation}

(v)$\{-A(k_{1}-i_{1}+M+1-j,j)\tilde{c}^{(j+1)}_{i_{1}-1}(k_{1}),-B(k_{2}-i_{2}+M+2-j,j)\tilde{c}^{(j+1)}_{i_{2}-2}(k_{2})\}$
gives
\begin{equation}
\begin{split}
&\rho(k_{1}-k_{2}-i_{1}+i_{2}-1,0)
+\zeta^{0,i_{2}-3}(k_{1}-i_{1}+M+1-j-k_{2},j) \\
&-\zeta^{1,i_{1}-2}(k_{2}-i_{2}+M+2-j-k_{1},j)
+\zeta^{i_{1}-2,i_{2}-3}(k_{1}-k_{2},0)\\
=&\zeta^{0,1}(k_{1}-k_{2}-i_{1}+i_{2}-1,0)
+\zeta^{0,i_{2}-3}(k_{1}-i_{1}+1-k_{2},0) \\
&+\zeta^{i_{1}-2,1}(k_{1}+i_{2}-2-k_{2},0)
+\zeta^{i_{1}-2,i_{2}-3}(k_{1}-k_{2},0) \\
&+\delta_{k_{1}-k_{2}-i_{1}+i_{2}-1,0}
-\delta_{k_{1}-k_{2}-i_{1}+i_{2}-1,-1}
+\delta_{k_{1}-i_{1}+1-k_{2},0} \\
&-\delta_{k_{1}-i_{1}+1-k_{2},-i_{2}+2}
+\delta_{k_{1}+i_{2}-2-k_{2},i_{1}-1}-\delta_{k_{1}+i_{2}-2-k_{2},0}
\\ =&\zeta^{0,i_{2}-1}(k_{1}-i_{1}+1-k_{2},0)
+\zeta^{i_{1}-2,0}(k_{1}+i_{2}-2-k_{2},0) \\
&+\zeta^{i_{1}-2,0}(k_{1}+i_{2}-1-k_{2},0)
+\zeta^{i_{1}-2,i_{2}-3}(k_{1}-k_{2},0) \\
&+\delta_{k_{1}+i_{2}-2-k_{2},i_{1}-2}
-\delta_{k_{1}+i_{2}-2-k_{2},-1}
+\delta_{k_{1}-k_{2}-i_{1}+i_{2}-1,0} \\
&-\delta_{k_{1}-k_{2}-i_{1}+i_{2}-1,-1}
+\delta_{k_{1}-i_{1}+1-k_{2},0}
-\delta_{k_{1}-i_{1}+1-k_{2},-i_{2}+2} \\
&+\delta_{k_{1}+i_{2}-2-k_{2},i_{1}-1}
-\delta_{k_{1}+i_{2}-2-k_{2},0} \\
=&\zeta^{0,i_{2}-1}(k_{1}-i_{1}+1-k_{2},0)
+\zeta^{i_{1}-2,i_{2}-1}(k_{1}-k_{2},0) \\
&+\delta_{k_{1}-k_{2},-i_{2}+2}
-\delta_{k_{1}-k_{2},-i_{2}+3+i_{1}-2}
+\delta_{k_{1}-k_{2},-i_{2}+1}  \\
&-\delta_{k_{1}-k_{2},-i_{2}+2+i_{1}-2}
+\delta_{k_{1}+i_{2}-2-k_{2},i_{1}-2}
-\delta_{k_{1}+i_{2}-2-k_{2},-1} \\
&+\delta_{k_{1}-k_{2}-i_{1}+i_{2}-1,0}
-\delta_{k_{1}-k_{2}-i_{1}+i_{2}-1,-1}
+\delta_{k_{1}-i_{1}+1-k_{2},0} \\
&-\delta_{k_{1}-i_{1}+1-k_{2},-i_{2}+2}
+\delta_{k_{1}+i_{2}-2-k_{2},i_{1}-1}-\delta_{k_{1}+i_{2}-2-k_{2},0}
\\ =&\zeta^{i_{1}-1,i_{2}-1}(k_{1}-k_{2},0)
+\delta_{k_{1}-k_{2},i_{1}-1-i_{2}+1}
-\delta_{k_{1}-k_{2},i_{1}-1} \\ &+\delta_{k_{1}-k_{2},-i_{2}+2}
-\delta_{k_{1}-k_{2},-i_{2}+3+i_{1}-2}
+\delta_{k_{1}-k_{2},-i_{2}+1} \\
&-\delta_{k_{1}-k_{2},-i_{2}+2+i_{1}-2}
+\delta_{k_{1}+i_{2}-2-k_{2},i_{1}-2}
-\delta_{k_{1}+i_{2}-2-k_{2},-1} \\
&+\delta_{k_{1}-k_{2}-i_{1}+i_{2}-1,0}
-\delta_{k_{1}-k_{2}-i_{1}+i_{2}-1,-1}
+\delta_{k_{1}-i_{1}+1-k_{2},0} \\
&-\delta_{k_{1}-i_{1}+1-k_{2},-i_{2}+2}
+\delta_{k_{1}+i_{2}-2-k_{2},i_{1}-1}-\delta_{k_{1}+i_{2}-2-k_{2},0}
\\ =&\zeta^{i_{1}-1,i_{2}-1}(k_{1}-k_{2},0) \\
\end{split}
\end{equation}

(vi)$\{-A(k_{1}-i_{1}+M+1-j,j)\tilde{c}^{(j+1)}_{i_{1}-1}(k_{1}),-A(k_{2}-i_{2}+M+1-j,j)\tilde{c}^{(j+1)}_{i_{2}-1}(k_{2})\}$
gives
\begin{equation}
\begin{split}
&\zeta^{0,0}(k_{1}-k_{2}-i_{1}+i_{2},0)
+\zeta^{0,i_{2}-2}(k_{1}-i_{1}+M+1-j-k_{2},j) \\
&-\zeta^{0,i_{1}-2}(k_{2}-i_{2}+M+1-j-k_{1},j)
+\zeta^{i_{1}-2,i_{2}-2}(k_{1}-k_{2},0)\\
=&\zeta^{0,0}(k_{1}-k_{2}-i_{1}+i_{2},0)
+\zeta^{0,i_{2}-2}(k_{1}-i_{1}+1-k_{2},0) \\
&+\zeta^{i_{1}-2,0}(k_{1}+i_{2}-1-k_{2},0)
+\zeta^{i_{1}-2,i_{2}-2}(k_{1}-k_{2},0)\\
&+\delta_{k_{1}-i_{1}+1-k_{2},0}-\delta_{k_{1}-i_{1}+1-k_{2},-i_{2}+1}
+\delta_{k_{1}+i_{2}-1-k_{2},i_{1}-1}-\delta_{k_{1}+i_{2}-1-k_{2},0}
\\ =&\zeta^{0,i_{2}-1}(k_{1}-i_{1}+1-k_{2},0)
+\zeta^{i_{1}-2,i_{2}-1}(k_{1}-k_{2},0)\\
&+\delta_{k_{1}-k_{2},-i_{2}+1}
-\delta_{k_{1}-k_{2},-i_{2}+1+i_{1}-1}\\
&+\delta_{k_{1}-i_{1}+1-k_{2},0}-\delta_{k_{1}-i_{1}+1-k_{2},-i_{2}+1}
+\delta_{k_{1}+i_{2}-1-k_{2},i_{1}-1}-\delta_{k_{1}+i_{2}-1-k_{2},0}
\\ =&\zeta^{i_{1}-1,i_{2}-1}(k_{1}-k_{2},0)
+\delta_{k_{1}-k_{2},i_{1}-1-i_{2}+1}
-\delta_{k_{1}-k_{2},i_{1}-1}\\ &+\delta_{k_{1}-k_{2},-i_{2}+1}
-\delta_{k_{1}-k_{2},-i_{2}+1+i_{1}-1}\\
&+\delta_{k_{1}-i_{1}+1-k_{2},0}-\delta_{k_{1}-i_{1}+1-k_{2},-i_{2}+1}
+\delta_{k_{1}+i_{2}-1-k_{2},i_{1}-1}-\delta_{k_{1}+i_{2}-1-k_{2},0}
\\ =&\zeta^{i_{1}-1,i_{2}-1}(k_{1}-k_{2},0) \\
\end{split}
\end{equation}

(vii)$\{-B(k_{1}-i_{1}+M+2-j,j)\tilde{c}^{(j+1)}_{i_{1}-2}(k_{1}),\tilde{c}^{(j+1)}_{i_{2}}(k_{2})\}$
gives
\begin{equation}
\begin{split}
&\zeta^{1,i_{2}-1}(k_{1}-i_{1}+M+2-j-k_{2},j)
+\zeta^{i_{1}-3,i_{2}-1}(k_{1}-k_{2},0) \\
=&\zeta^{1,i_{2}-1}(k_{1}-k_{2}-i_{1}+2,0)
+\zeta^{i_{1}-3,i_{2}-1}(k_{1}-k_{2},0)
+\delta_{k_{1}-k_{2}-i_{1}+2,0}-\delta_{k_{1}-k_{2}-i_{1}+2,-i_{2}}
\\ =&\zeta^{0,i_{2}-1}(k_{1}-k_{2}-i_{1}+2,0)
+\zeta^{0,i_{2}-1}(k_{1}-k_{2}-i_{1}+1,0) \\
&+\zeta^{i_{1}-3,i_{2}-1}(k_{1}-k_{2},0)
+\delta_{k_{1}-k_{2}-i_{1}+2,1} \\
&-\delta_{k_{1}-k_{2}-i_{1}+2,-i_{2}+1}
+\delta_{k_{1}-k_{2}-i_{1}+2,0}-\delta_{k_{1}-k_{2}-i_{1}+2,-i_{2}}
\\ =&\zeta^{i_{1}-1,i_{2}-1}(k_{1}-k_{2},0)
+\delta_{k_{1}-k_{2},i_{1}-2-i_{2}+1}
-\delta_{k_{1}-k_{2},i_{1}-2} \\
&+\delta_{k_{1}-k_{2},i_{1}-1-i_{2}+1}
-\delta_{k_{1}-k_{2},i_{1}-1} +\delta_{k_{1}-k_{2}-i_{1}+2,1} \\
&-\delta_{k_{1}-k_{2}-i_{1}+2,-i_{2}+1}
+\delta_{k_{1}-k_{2}-i_{1}+2,0}
-\delta_{k_{1}-k_{2}-i_{1}+2,-i_{2}} \\
=&\zeta^{i_{1}-1,i_{2}-1}(k_{1}-k_{2},0)
+\delta_{k_{1}-i_{1},k_{2}-i_{2}}
-\delta_{k_{1}-i_{1}+2,k_{2}-i_{2}} \\
\end{split}
\end{equation}

(viii)$\{-B(k_{1}-i_{1}+M+2-j,j)\tilde{c}^{(j+1)}_{i_{1}-2}(k_{1}),-A(k_{2}-i_{2}+M+1-j,j)\tilde{c}^{(j+1)}_{i_{2}-1}(k_{2})\}$
gives
\begin{equation}
\begin{split}
&-\rho(k_{2}-k_{1}-i_{2}+i_{1}-1,0)
+\zeta^{1,i_{2}-2}(k_{1}-i_{1}+M+2-j-k_{2},j) \\
&-\zeta^{0,i_{1}-3}(k_{2}-i_{2}+M+1-j-k_{1},j)
+\zeta^{i_{1}-3,i_{2}-2}(k_{1}-k_{2},0) \\
=&\zeta^{1,0}(k_{1}-k_{2}+i_{2}-i_{1}+1,0)
+\zeta^{1,i_{2}-2}(k_{1}-i_{1}+2-k_{2},0) \\
&+\zeta^{i_{1}-3,0}(k_{1}+i_{2}-1-k_{2},0)
+\zeta^{i_{1}-3,i_{2}-2}(k_{1}-k_{2},0) \\
&+\delta_{k_{2}-k_{1}-i_{2}+i_{1}-1,-1}
-\delta_{k_{2}-k_{1}-i_{2}+i_{1}-1,0}
+\delta_{k_{1}-i_{1}+2-k_{2},0} \\
&-\delta_{k_{1}-i_{1}+2-k_{2},-i_{2}+1}
+\delta_{k_{1}+i_{2}-1-k_{2},i_{1}-2}-\delta_{k_{1}+i_{2}-1-k_{2},0}
\\ =&\zeta^{i_{1}-1,0}(k_{1}-k_{2}+i_{2}-1,0)
+\zeta^{0,i_{2}-2}(k_{1}-i_{1}+2-k_{2},0) \\
&+\zeta^{0,i_{2}-2}(k_{1}-i_{1}+1-k_{2},0)
+\zeta^{i_{1}-3,i_{2}-2}(k_{1}-k_{2},0) \\
&+\delta_{k_{1}-i_{1}+2-k_{2},1}
-\delta_{k_{1}-i_{1}+2-k_{2},-i_{2}+2}
+\delta_{k_{2}-k_{1}-i_{2}+i_{1}-1,-1} \\
&-\delta_{k_{2}-k_{1}-i_{2}+i_{1}-1,0}
+\delta_{k_{1}-i_{1}+2-k_{2},0}
-\delta_{k_{1}-i_{1}+2-k_{2},-i_{2}+1} \\
&+\delta_{k_{1}+i_{2}-1-k_{2},i_{1}-2}-\delta_{k_{1}+i_{2}-1-k_{2},0}
\\ =&\zeta^{i_{1}-1,0}(k_{1}-k_{2}+i_{2}-1,0)
+\zeta^{i_{1}-1,i_{2}-2}(k_{1}-k_{2},0) \\
&+\delta_{k_{1}-k_{2},i_{1}-1-i_{2}+2}
-\delta_{k_{1}-k_{2},i_{1}-1}
+\delta_{k_{1}-k_{2},i_{1}-2-i_{2}+2} \\
&-\delta_{k_{1}-k_{2},i_{1}-2} +\delta_{k_{1}-i_{1}+2-k_{2},1}
-\delta_{k_{1}-i_{1}+2-k_{2},-i_{2}+2} \\
&+\delta_{k_{2}-k_{1}-i_{2}+i_{1}-1,-1}
-\delta_{k_{2}-k_{1}-i_{2}+i_{1}-1,0}
+\delta_{k_{1}-i_{1}+2-k_{2},0} \\
&-\delta_{k_{1}-i_{1}+2-k_{2},-i_{2}+1}
+\delta_{k_{1}+i_{2}-1-k_{2},i_{1}-2}-\delta_{k_{1}+i_{2}-1-k_{2},0}
\\ =&\zeta^{i_{1}-1,i_{2}-1}(k_{1}-k_{2},0)
+\delta_{k_{1}-k_{2},-i_{2}+1}
-\delta_{k_{1}-k_{2},-i_{2}+2+i_{1}-1} \\
&+\delta_{k_{1}-k_{2},i_{1}-1-i_{2}+2}
-\delta_{k_{1}-k_{2},i_{1}-1}
+\delta_{k_{1}-k_{2},i_{1}-2-i_{2}+2} \\
&-\delta_{k_{1}-k_{2},i_{1}-2} +\delta_{k_{1}-i_{1}+2-k_{2},1}
-\delta_{k_{1}-i_{1}+2-k_{2},-i_{2}+2} \\
&+\delta_{k_{2}-k_{1}-i_{2}+i_{1}-1,-1}
-\delta_{k_{2}-k_{1}-i_{2}+i_{1}-1,0}
+\delta_{k_{1}-i_{1}+2-k_{2},0} \\
&-\delta_{k_{1}-i_{1}+2-k_{2},-i_{2}+1}
+\delta_{k_{1}+i_{2}-1-k_{2},i_{1}-2}-\delta_{k_{1}+i_{2}-1-k_{2},0}
\\ =&\zeta^{i_{1}-1,i_{2}-1}(k_{1}-k_{2},0)
+\delta_{k_{1}-i_{1},k_{2}-i_{2}}-\delta_{k_{1}-i_{1}+1,k_{2}-i_{2}}\\
\end{split}
\end{equation}

(ix)$\{-B(k_{1}-i_{1}+M+2-j,j)\tilde{c}^{(j+1)}_{i_{1}-2}(k_{1}),-B(k_{2}-i_{2}+M+2-j,j)\tilde{c}^{(j+1)}_{i_{2}-2}(k_{2})\}$
gives
\begin{equation}
\begin{split}
&\phi(k_{1}-k_{2}-i_{1}+i_{2},0)
+\zeta^{1,i_{2}-3}(k_{1}-i_{1}+M+2-j-k_{2},j) \\
&-\zeta^{1,i_{1}-3}(k_{2}-i_{2}+M+2-j-k_{1},j)
+\zeta^{i_{1}-3,i_{2}-3}(k_{1}-k_{2},0)\\
=&\zeta^{1,1}(k_{1}-k_{2}-i_{1}+i_{2},0)
+\zeta^{1,i_{2}-3}(k_{1}-i_{1}+2-k_{2},0) \\
&+\zeta^{i_{1}-3,1}(k_{1}+i_{2}-2-k_{2},0)
+\zeta^{i_{1}-3,i_{2}-3}(k_{1}-k_{2},0)\\
&+\delta_{k_{1}-i_{1}+2-k_{2},0}-\delta_{k_{1}-i_{1}+2-k_{2},-i_{2}+2}
+\delta_{k_{1}+i_{2}-2-k_{2},i_{1}-3}-\delta_{k_{1}+i_{2}-2-k_{2},0}
\\ =&\zeta^{0,1}(k_{1}-k_{2}-i_{1}+i_{2},0)
+\zeta^{0,1}(k_{1}-k_{2}-i_{1}+i_{2}-1,0) \\
&+\zeta^{0,i_{2}-3}(k_{1}-i_{1}+2-k_{2},0)
+\zeta^{0,i_{2}-3}(k_{1}-i_{1}+1-k_{2},0) \\
&+\zeta^{i_{1}-3,0}(k_{1}+i_{2}-2-k_{2},0)
+\zeta^{i_{1}-3,0}(k_{1}+i_{2}-1-k_{2},0) \\
&+\zeta^{i_{1}-3,i_{2}-3}(k_{1}-k_{2},0)
+\delta_{k_{1}-k_{2}-i_{1}+i_{2},1}-\delta_{k_{1}-k_{2}-i_{1}+i_{2},-1}
\\
&+\delta_{k_{1}-i_{1}+2-k_{2},1}-\delta_{k_{1}-i_{1}+2-k_{2},-i_{2}+3}
+\delta_{k_{1}+i_{2}-2-k_{2},i_{1}-3} \\
&-\delta_{k_{1}+i_{2}-2-k_{2},-1}
+\delta_{k_{1}-i_{1}+2-k_{2},0}-\delta_{k_{1}-i_{1}+2-k_{2},-i_{2}+2}
\\
&+\delta_{k_{1}+i_{2}-2-k_{2},i_{1}-3}-\delta_{k_{1}+i_{2}-2-k_{2},0}
\\ =&\zeta^{i_{1}-1,i_{2}-1}(k_{1}-k_{2},0)
+\delta_{k_{1}-k_{2},i_{1}-1-i_{2}+1}-\delta_{k_{1}-k_{2},i_{1}-1}
\\
&+\delta_{k_{1}-k_{2},i_{1}-1-i_{2}+1}-\delta_{k_{1}-k_{2},i_{1}-2}
+\delta_{k_{1}-k_{2},-i_{2}+1} \\
&-\delta_{k_{1}-k_{2},-i_{2}+2+i_{1}-3}
+\delta_{k_{1}-k_{2},-i_{2}+2}-\delta_{k_{1}-k_{2},-i_{2}+3+i_{1}-3}
\\
&+\delta_{k_{1}-k_{2}-i_{1}+i_{2},1}-\delta_{k_{1}-k_{2}-i_{1}+i_{2},-1}
+\delta_{k_{1}-i_{1}+2-k_{2},1} \\
&-\delta_{k_{1}-i_{1}+2-k_{2},-i_{2}+3}
+\delta_{k_{1}+i_{2}-2-k_{2},i_{1}-3}-\delta_{k_{1}+i_{2}-2-k_{2},-1}
\\
&+\delta_{k_{1}-i_{1}+2-k_{2},0}-\delta_{k_{1}-i_{1}+2-k_{2},-i_{2}+2}
+\delta_{k_{1}+i_{2}-2-k_{2},i_{1}-3} \\
&-\delta_{k_{1}+i_{2}-2-k_{2},0} \\
=&\zeta^{i_{1}-1,i_{2}-1}(k_{1}-k_{2},0)
\end{split}
\end{equation}

Except for (iv),(vii) and (viii), notice that all of these
coefficients are identically
$\zeta^{i_{1}-1,i_{2}-1}(k_{1}-k_{2},0)$. Let's rewrite the extra
terms that appear in (iv),(vii),(viii). Remember that in the
formulae above, $\delta_{a,b}$ is $1$ iff $a$ and $b$ are
equivalent modulo $N$, so each of them actually represents an
infinite sum of $\delta$'s if $a,b$ are considered to be in
$\mathbb{Z}$. The totality of the excess terms is the sum of the
expression below over $l\in\mathbb{Z}$:

\begin{equation}\label{eq:A1}
\begin{split}
-\delta_{k_{1}-i_{1},k_{2}+lN-i_{2}}
&A(k_{1}-i_{1}+M+1-j,j)\tilde{c}^{(j+1)}_{i_{1}-1}(k_{1})\tilde{c}^{(j+1)}_{i_{2}}(k_{2})
\\
+\delta_{k_{1}-i_{1}+1,k_{2}+lN-i_{2}}
&A(k_{1}-i_{1}+M+1-j,j)\tilde{c}^{(j+1)}_{i_{1}-1}(k_{1})\tilde{c}^{(j+1)}_{i_{2}}(k_{2})
\\
-\delta_{k_{1}-i_{1},k_{2}+lN-i_{2}}
&B(k_{1}-i_{1}+M+2-j,j)\tilde{c}^{(j+1)}_{i_{1}-2}(k_{1})\tilde{c}^{(j+1)}_{i_{2}}(k_{2})
\\
+\delta_{k_{1}-i_{1}+2,k_{2}+lN-i_{2}}
&B(k_{1}-i_{1}+M+2-j,j)\tilde{c}^{(j+1)}_{i_{1}-2}(k_{1})\tilde{c}^{(j+1)}_{i_{2}}(k_{2})
\\
+\delta_{k_{1}-i_{1},k_{2}+lN-i_{2}}
&B(k_{1}-i_{1}+M+2-j,j)\tilde{c}^{(j+1)}_{i_{1}-2}(k_{1})\\
&A(k_{2}-i_{2}+M+1-j,j)\tilde{c}^{(j+1)}_{i_{2}-1}(k_{2})
\\
-\delta_{k_{1}-i_{1}+1,k_{2}+lN-i_{2}}
&B(k_{1}-i_{1}+M+2-j,j)\tilde{c}^{(j+1)}_{i_{1}-2}(k_{1})\\
&A(k_{2}-i_{2}+M+1-j,j)\tilde{c}^{(j+1)}_{i_{2}-1}(k_{2})
\\
\end{split}
\end{equation}
Next we look at

\begin{equation}
\begin{split}
\{&-A(k_{1}-i_{1}+M+1-j,j)\tilde{c}^{(j+1)}_{i_{1}-1}(k_{1}), \\
 &-A(k_{2}-i_{2}+M+1-j,j)\tilde{c}^{(j+1)}_{i_{2}-1}(k_{2})\} \\
\end{split}
\end{equation}
When this bracket is expanded by Leibniz rule, there is an
$\{A,A\}$ term, which, by $\eqref{eq:bracket}$, will result in a
$B$ if the $A$'s are horizontal neighbors. Since this is not a
product term, we haven't taken it into account yet. Notice that,
within one period this happens only for $\{A(n,m),A(n-1,m)\}$ and
$\{A(n,m),A(n+1,m)\}$. For the full infinite set of indices, in
our case, a nonproduct term will arise when
$k_{1}-i_{1}+M+1-j=k_{2}-i_{2}+M+1-j+lN+1$, or
$k_{1}-i_{1}+M+1-j=k_{2}-i_{2}+M+1-j+lN-1$, for some integer $l$.
Simplifying, the conditions become $k_{1}-i_{1}=k_{2}-i_{2}+lN+1$
or $k_{1}-i_{1}=k_{2}-i_{2}+lN-1$ for some integer $l$. And in
these cases, what one gets for the nonproduct term is

\begin{equation} \label{eq:A2}
\begin{split}
&\delta_{k_{1}-i_{1},k_{2}-i_{2}+lN-1}B(k_{2}-i_{2}+M+1-j,j)\tilde{c}_{i_{1}-1}^{(j+1)}(k_{1})\tilde{c}_{i_{2}-1}^{(j+1)}(k_{2})
\\
&-\delta_{k_{1}-i_{1},k_{2}-i_{2}+lN+1}B(k_{2}-i_{2}+M+2-j,j)\tilde{c}_{i_{1}-1}^{(j+1)}(k_{1})\tilde{c}_{i_{2}-1}^{(j+1)}(k_{2})
\\
\end{split}
\end{equation}
Next, we look at
$f(\tilde{c}_{i_{1}}^{(j)}(k_{1}),\tilde{c}_{i_{2}}^{(j)}(k_{2}+lN))$,
and compare this to the sum of $9$ $f$ terms obtained from the
expansion of $\eqref{eq:expand1}$.

\begin{equation} \label{eq:f1}
\begin{split}
&f(\tilde{c}_{i_{1}}^{(j)}(k_{1}),\tilde{c}_{i_{2}}^{(j)}(k_{2}+lN))
\\ =& (\delta_{k_{2}+lN\leq k_{1}}\delta_{k_{2}+lN-i_{2}\leq
k_{1}-i_{1}}-\delta_{k_{2}+lN\geq k_{1}}\delta_{k_{2}+lN-i_{2}\geq
k_{1}-i_{1}})\\
&\tilde{c}_{k_{1}-k_{2}-lN+i_{2}}^{(j)}(k_{1})\tilde{c}_{k_{2}+lN-k_{1}+i_{1}}^{(j)}(k_{2}+lN)
\\
=& (\delta_{k_{2}+lN\leq k_{1}}\delta_{k_{2}+lN-i_{2}\leq
k_{1}-i_{1}}-\delta_{k_{2}+lN\geq k_{1}}\delta_{k_{2}+lN-i_{2}\geq
k_{1}-i_{1}})
\\
&(\tilde{c}_{k_{1}-k_{2}-lN+i_{2}}^{(j+1)}(k_{1})-A(k_{2}+lN-i_{2}+M+1-j,j)\tilde{c}_{k_{1}-k_{2}-lN+i_{2}-1}^{(j+1)}(k_{1})
\\
&-B(k_{2}+lN-i_{2}+M+2-j,j)\tilde{c}_{k_{1}-k_{2}-lN+i_{2}-2}^{(j+1)}(k_{1}))
\\
&(\tilde{c}_{k_{2}+lN-k_{1}+i_{1}}^{(j+1)}(k_{2}+lN)-A(k_{1}-i_{1}+M+1-j,j)\tilde{c}_{k_{2}+lN-k_{1}+i_{1}-1}^{(j+1)}(k_{2}+lN)
\\
&-B(k_{1}-i_{1}+M+2-j,j)\tilde{c}_{k_{2}+lN-k_{1}+i_{1}-2}^{(j+1)}(k_{2}+lN))
\\
\end{split}
\end{equation}
From $\eqref{eq:expand1}$ we get the following $9$ $f$ terms for
each integer $l$.

\begin{equation} \label{eq:f2}
\begin{split}
&\{\tilde{c}^{(j+1)}_{i_{1}}(k_{1})-A(k_{1}-i_{1}+M+1-j,j)\tilde{c}^{(j+1)}_{i_{1}-1}(k_{1})
\\
&-B(k_{1}-i_{1}+M+2-j,j)\tilde{c}^{(j+1)}_{i_{1}-2}(k_{1}),
\\
&\tilde{c}^{(j+1)}_{i_{2}}(k_{2})-A(k_{2}-i_{2}+M+1-j,j)\tilde{c}^{(j+1)}_{i_{2}-1}(k_{2})
\\
&-B(k_{2}-i_{2}+M+2-j,j)\tilde{c}^{(j+1)}_{i_{2}-2}(k_{2})\}
\\
\rightarrow &
f(\tilde{c}_{i_{1}}^{(j+1)}(k_{1}),\tilde{c}_{i_{2}}^{(j+1)}(k_{2}+lN))
\\
&-A(k_{2}-i_{2}+M+1-j,j)f(\tilde{c}_{i_{1}}^{(j+1)}(k_{1}),\tilde{c}_{i_{2}-1}^{(j+1)}(k_{2}+lN))
\\
&-B(k_{2}-i_{2}+M+2-j,j)f(\tilde{c}_{i_{1}}^{(j+1)}(k_{1}),\tilde{c}_{i_{2}-2}^{(j+1)}(k_{2}+lN))
\\
&-A(k_{1}-i_{1}+M+1-j,j)f(\tilde{c}_{i_{1}-1}^{(j+1)}(k_{1}),\tilde{c}_{i_{2}}^{(j+1)}(k_{2}+lN))
\\
&+A(k_{1}-i_{1}+M+1-j,j)A(k_{2}-i_{2}+M+1-j,j) \\
&f(\tilde{c}_{i_{1}-1}^{(j+1)}(k_{1}),\tilde{c}_{i_{2}-1}^{(j+1)}(k_{2}+lN))
\\
&+A(k_{1}-i_{1}+M+1-j,j)B(k_{2}-i_{2}+M+2-j,j)\\
&f(\tilde{c}_{i_{1}-1}^{(j+1)}(k_{1}),\tilde{c}_{i_{2}-2}^{(j+1)}(k_{2}+lN))
\\
&-B(k_{2}-i_{2}+M+2-j,j)f(\tilde{c}_{i_{1}-2}^{(j+1)}(k_{1}),\tilde{c}_{i_{2}}^{(j+1)}(k_{2}+lN))
\\
&+B(k_{1}-i_{1}+M+2-j,j)A(k_{2}-i_{2}+M+1-j,j)\\
&f(\tilde{c}_{i_{1}-2}^{(j+1)}(k_{1}),\tilde{c}_{i_{2}-1}^{(j+1)}(k_{2}+lN))
\\
&+B(k_{1}-i_{1}+M+2-j,j)B(k_{2}-i_{2}+M+2-j,j)\\
&f(\tilde{c}_{i_{1}-2}^{(j+1)}(k_{1}),\tilde{c}_{i_{2}-2}^{(j+1)}(k_{2}+lN))
\\
\end{split}
\end{equation}
If the $f$'s in the last expression are expanded, one can
calculate $\eqref{eq:f2}$ minus $\eqref{eq:f1}$. Most terms
cancel, but some boundary terms remain. The difference turns out
to be:

\begin{equation}\label{eq:longdiff}
\begin{split}
&\delta_{k_{2}+lN\geq
k_{1}}\delta_{k_{2}+lN-i_{2}+1,k_{1}-i_{1}}A(k_{2}-i_{2}+M+1-j,j)\tilde{c}_{i_{1}}^{(j+1)}(k_{1})\tilde{c}_{i_{2}-1}^{(j+1)}(k_{2})
\\
+&\delta_{k_{2}+lN\leq
k_{1}}\delta_{k_{2}+lN-i_{2},k_{1}-i_{1}}A(k_{2}-i_{2}+M+1-j,j)\tilde{c}_{i_{1}-1}^{(j+1)}(k_{1})\tilde{c}_{i_{2}}^{(j+1)}(k_{2})
\\
+&\delta_{k_{2}+lN\geq
k_{1}}\delta_{k_{2}+lN-i_{2}+1,k_{1}-i_{1}}B(k_{2}-i_{2}+M+2-j,j)\tilde{c}_{i_{1}-1}^{(j+1)}(k_{1})\tilde{c}_{i_{2}-1}^{(j+1)}(k_{2})
\\
+&\delta_{k_{2}+lN\geq
k_{1}}\delta_{k_{2}+lN-i_{2}+2,k_{1}-i_{1}}B(k_{2}-i_{2}+M+2-j,j)\tilde{c}_{i_{1}}^{(j+1)}(k_{1})\tilde{c}_{i_{2}-2}^{(j+1)}(k_{2})
\\
+&\delta_{k_{2}+lN\leq
k_{1}}\delta_{k_{2}+lN-i_{2},k_{1}-i_{1}}B(k_{2}-i_{2}+M+2-j,j)\tilde{c}_{i_{1}-2}^{(j+1)}(k_{1})\tilde{c}_{i_{2}}^{(j+1)}(k_{2})
\\
+&\delta_{k_{2}+lN\leq
k_{1}}\delta_{k_{2}+lN-i_{2}+1,k_{1}-i_{1}}B(k_{2}-i_{2}+M+2-j,j)\tilde{c}_{i_{1}-1}^{(j+1)}(k_{1})\tilde{c}_{i_{2}-1}^{(j+1)}(k_{2})
\\
-&\delta_{k_{2}+lN\geq
k_{1}}\delta_{k_{2}+lN-i_{2}+1,k_{1}-i_{1}}A(k_{1}-i_{1}+M+1-j,j)\\
&B(k_{2}-i_{2}+M+2-j,j)\tilde{c}_{i_{1}-1}^{(j+1)}(k_{1})\tilde{c}_{i_{2}-2}^{(j+1)}(k_{2})
\\
-&\delta_{k_{2}+lN\leq
k_{1}}\delta_{k_{2}+lN-i_{2},k_{1}-i_{1}}A(k_{1}-i_{1}+M+1-j,j)\\
&B(k_{2}-i_{2}+M+2-j,j)\tilde{c}_{i_{1}-2}^{(j+1)}(k_{1})\tilde{c}_{i_{2}-1}^{(j+1)}(k_{2})
\\
-&\delta_{k_{2}+lN\leq
k_{1}}\delta_{k_{2}+lN-i_{2},k_{1}-i_{1}+1}B(k_{1}-i_{1}+M+2-j,j)\tilde{c}_{i_{1}-1}^{(j+1)}(k_{1})\tilde{c}_{i_{2}-1}^{(j+1)}(k_{2})
\\
-&\delta_{k_{2}+lN\leq
k_{1}}\delta_{k_{2}+lN-i_{2},k_{1}-i_{1}+2}B(k_{1}-i_{1}+M+2-j,j)\tilde{c}_{i_{1}-2}^{(j+1)}(k_{1})\tilde{c}_{i_{2}}^{(j+1)}(k_{2})
\\
-&\delta_{k_{2}+lN\geq
k_{1}}\delta_{k_{2}+lN-i_{2},k_{1}-i_{1}}B(k_{1}-i_{1}+M+2-j,j)\tilde{c}_{i_{1}}^{(j+1)}(k_{1})\tilde{c}_{i_{2}-2}^{(j+1)}(k_{2})
\\
-&\delta_{k_{2}+lN\geq
k_{1}}\delta_{k_{2}+lN-i_{2},k_{1}-i_{1}+1}B(k_{1}-i_{1}+M+2-j,j)\tilde{c}_{i_{1}-1}^{(j+1)}(k_{1})\tilde{c}_{i_{2}-1}^{(j+1)}(k_{2})
\\
+&\delta_{k_{2}+lN\leq
k_{1}}\delta_{k_{2}+lN-i_{2}+1,k_{1}-i_{1}+2}B(k_{1}-i_{1}+M+2-j,j)\\
&A(k_{2}-i_{2}+M+1-j,j)\tilde{c}_{i_{1}-2}^{(j+1)}(k_{1})\tilde{c}_{i_{2}-1}^{(j+1)}(k_{2})
\\
+&\delta_{k_{2}+lN\geq
k_{1}}\delta_{k_{2}+lN-i_{2},k_{1}-i_{1}}B(k_{1}-i_{1}+M+2-j,j)\\
&A(k_{2}-i_{2}+M+1-j,j)
\tilde{c}_{i_{1}-1}^{(j+1)}(k_{1})\tilde{c}_{i_{2}-2}^{(j+1)}(k_{2})
\\
-&\delta_{k_{2}+lN\leq
k_{1}}\delta_{k_{2}+lN-i_{2},k_{1}-i_{1}+1}A(k_{1}-i_{1}+M+1-j,j)\tilde{c}_{i_{1}-1}^{(j+1)}(k_{1})\tilde{c}_{i_{2}}^{(j+1)}(k_{2})
\\
-&\delta_{k_{2}+lN\geq
k_{1}}\delta_{k_{2}+lN-i_{2},k_{1}-i_{1}}A(k_{1}-i_{1}+M+1-j,j)\tilde{c}_{i_{1}}^{(j+1)}(k_{1})\tilde{c}_{i_{2}-1}^{(j+1)}(k_{2})
\\
\end{split}
\end{equation}
Notice that, by our assumption $i_{1}\geq i_{2}+3$, all of the
terms above that start with $\delta_{k_{2}\geq k_{1}}$, namely
half of them, drop out. For instance, look at the first term,
which begins with $\delta_{k_{2}\geq
k_{1}}\delta_{k_{2}-i_{2}+1,k_{1}-i_{1}}$. If
$k_{2}-i_{2}+1=k_{1}-i_{1}$, then $k_{2}-k_{1}=i_{2}-i_{1}-1<0$.
Thus $k_{2}\geq k_{1}$ cannot be satisfied. Removing those,
$\eqref{eq:longdiff}$ becomes:

\begin{equation}\label{eq:A3}
\begin{split}
&\delta_{k_{2}+lN-i_{2},k_{1}-i_{1}}A(k_{2}-i_{2}+M+1-j,j)\tilde{c}_{i_{1}-1}^{(j+1)}(k_{1})\tilde{c}_{i_{2}}^{(j+1)}(k_{2})
\\
+&\delta_{k_{2}+lN-i_{2},k_{1}-i_{1}}B(k_{2}-i_{2}+M+2-j,j)\tilde{c}_{i_{1}-2}^{(j+1)}(k_{1})\tilde{c}_{i_{2}}^{(j+1)}(k_{2})
\\
+&\delta_{k_{2}+lN-i_{2}+1,k_{1}-i_{1}}B(k_{2}-i_{2}+M+2-j,j)\tilde{c}_{i_{1}-1}^{(j+1)}(k_{1})\tilde{c}_{i_{2}-1}^{(j+1)}(k_{2})
\\
-&\delta_{k_{2}+lN-i_{2},k_{1}-i_{1}}A(k_{1}-i_{1}+M+1-j,j)\\
&B(k_{2}-i_{2}+M+2-j,j)\tilde{c}_{i_{1}-2}^{(j+1)}(k_{1})\tilde{c}_{i_{2}-1}^{(j+1)}(k_{2})
\\
-&\delta_{k_{2}+lN-i_{2},k_{1}-i_{1}+1}B(k_{1}-i_{1}+M+2-j,j)\tilde{c}_{i_{1}-1}^{(j+1)}(k_{1})\tilde{c}_{i_{2}-1}^{(j+1)}(k_{2})
\\
-&\delta_{k_{2}+lN-i_{2},k_{1}-i_{1}+2}B(k_{1}-i_{1}+M+2-j,j)\tilde{c}_{i_{1}-2}^{(j+1)}(k_{1})\tilde{c}_{i_{2}}^{(j+1)}(k_{2})
\\
+&\delta_{k_{2}+lN-i_{2}+1,k_{1}-i_{1}+2}B(k_{1}-i_{1}+M+2-j,j)\\
&A(k_{2}-i_{2}+M+1-j,j)\tilde{c}_{i_{1}-2}^{(j+1)}(k_{1})\tilde{c}_{i_{2}-1}^{(j+1)}(k_{2})
\\
-&\delta_{k_{2}+lN-i_{2},k_{1}-i_{1}+1}A(k_{1}-i_{1}+M+1-j,j)\tilde{c}_{i_{1}-1}^{(j+1)}(k_{1})\tilde{c}_{i_{2}}^{(j+1)}(k_{2})
\\
\end{split}
\end{equation}
As a final step, check that
$\eqref{eq:A1}+\eqref{eq:A2}+\eqref{eq:A3}=0$, This concludes the
proof. $\Box$

\clearpage

\noindent \textbf{APPENDIX 2: Examples of Toroidal Pipe Diagrams}

We give some examples for section \ref{sec:combinatorial} in this
appendix. The first example shows an instance of the decomposition
implied in corollary \ref{cor:tpd}, whereas the second example
shows that its hypothesis is not vacuous.

\begin{figure}
\begin{center}
\epsfig{file=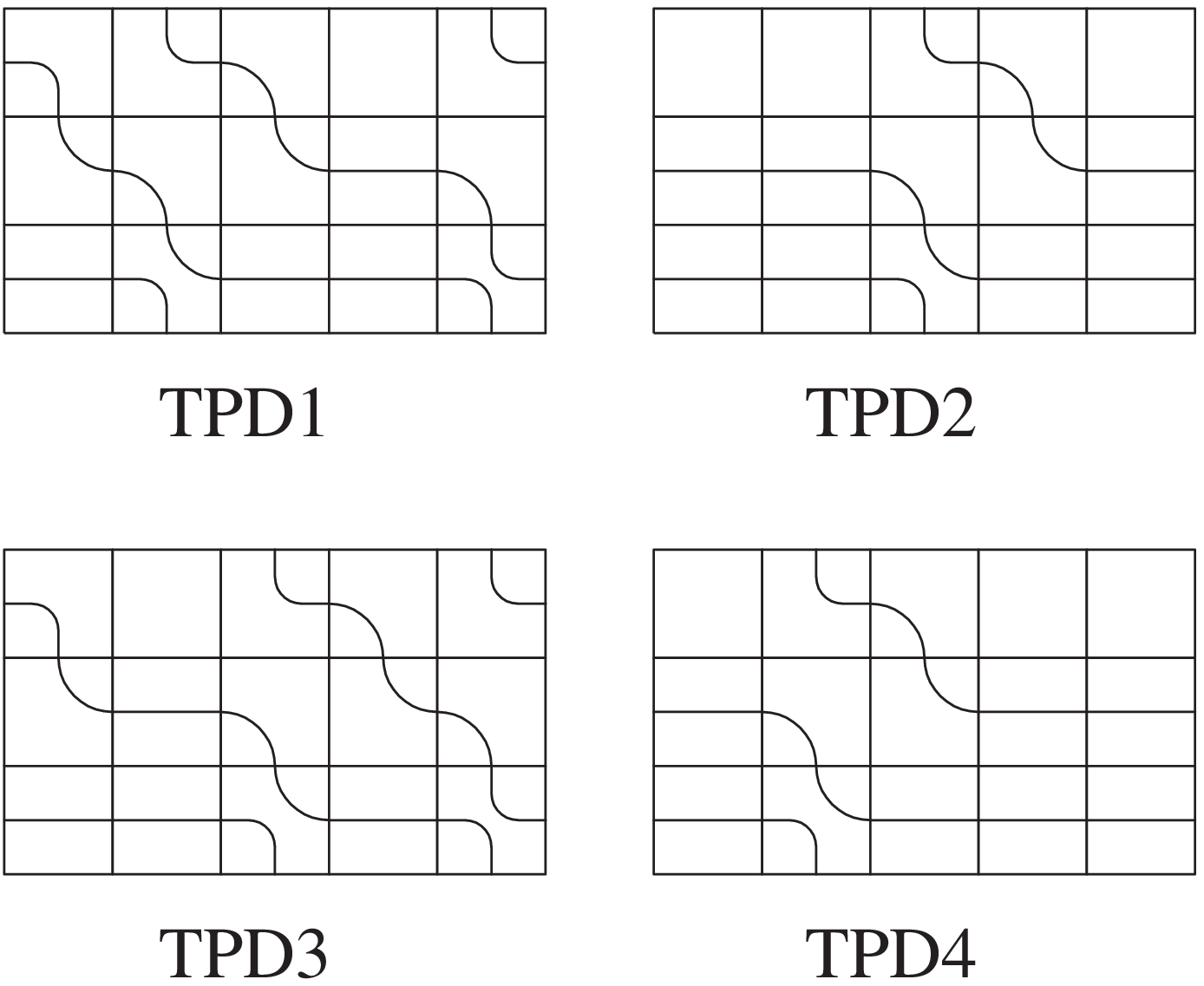,height=8.1cm,width=9.9cm}
\caption{}\label{fig:1}
\end{center}
\end{figure}

In figure \ref{fig:1}, $<\mathcal{TPD}_{1},\mathcal{TPD}_{2}>=-1$.
There is only one other decomposition of
$\mathcal{TPD}_{1}\times\mathcal{TPD}_{2}$ into two toroidal pipe
diagrams of respective degrees 4  and 7 : $\mathcal{TPD}_{3}$ and
$\mathcal{TPD}_{4}$ of the same figure.
$<\mathcal{TPD}_{3},\mathcal{TPD}_{4}>=1$.

\begin{figure}
\begin{center}
\epsfig{file=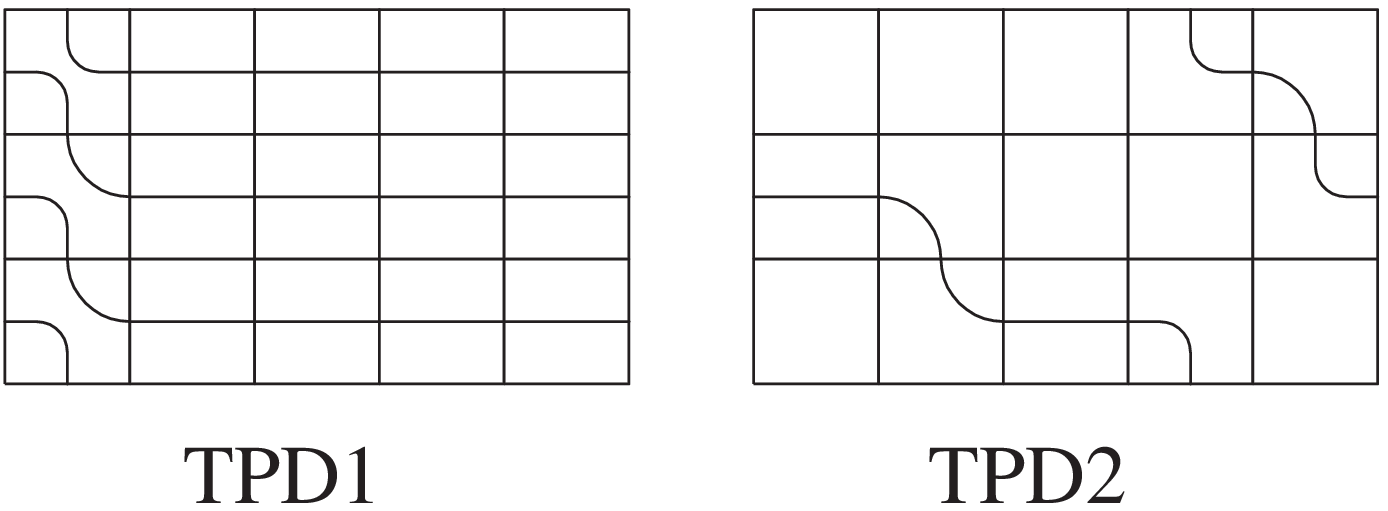,height=3.6cm,width=9.9cm} \caption{}
\label{fig:2}
\end{center}
\end{figure}

In figure \ref{fig:2}, $<\mathcal{TPD}_{1},\mathcal{TPD}_{2}>=0$,
and this is the only decomposition of
$\mathcal{TPD}_{1}\times\mathcal{TPD}_{2}$ into two toroidal pipe
diagrams of respective degrees 12 and 2 .

\clearpage


\begin{thebibliography}{50}

\bibitem[Adl] {Adl} Adler, M. {\em On a trace functional for formal
pseudo-differential operators and the symplectic structure of the
Korteweg-de Vries equation} Invent. Math. , {\bf 50} No:3
(1979), 219---248

\bibitem[A-vM] {A-vM} Adler, M. , van Moerbeke, P. {\em Generalized
orthogonal polynomials, discrete KP and Riemann-Hilbert problems} Comm.
in Math. Phys. , {\bf 207} (1999), 589---620

\bibitem[Akh] {Akh} Akhiezer, N. I. {\em A continuous analog of
orthogonal polynomials on a system of integrals} Soviet Math. Dokl. , {\bf
2} (1961) 1409---1412

\bibitem[Arn] {Arn} Arnold, V. I. {\em Mathematical Methods of Classical
Mechanics} Graduate Texts in Mathematics, {\bf 60} Springer-Verlag
(1978)

\bibitem[A-G]{A-G} Arnold, V. I. , Givental, A. B. {\em Symplectic
Geometry} Encyclopedia of Mathematical Sciences, Dynamical Systems IV
(1985), 5---131

\bibitem[B-C]{B-C} Burchnall, J. L. , Chaundy, T. W. {\em Commutative
ordinary differential operators} Proc. London Math. Soc. Series 2, {\bf
21} (1923) 420---440

\bibitem[C-S]{C-S} Chas, M. , Sullivan, D. {\em String Topology} preprint
arxiv:math.GT/9911159

\bibitem[Dam]{Dam} Damianou, P. A. {\em Multiple Hamiltonian structures
for Toda type systems.} J. of Math. Phys. {\bf 35} (1994), 5511---5541

\bibitem[Dic]{Dic} Dickey, L. A. {\em Lectures on Classical $\mathcal{W}$
Algebras} Acta Applicandae Math.  {\bf 47} (1997), 243---321

\bibitem[D-S]{D-S} Drinfeld, V. G. , Sokolov, V. V. {\em Lie algebras and
equations of Korteweg-de Vries type} J. Sov. Math. {\bf 30}
(1985), 1975---2036

\bibitem[Dub]{Dub} Dubrovin, B. A. {\em Theta functions and nonlinear
equations} Russian Math. Surveys {\bf 36} (1981), 11---92

\bibitem[DKN]{DKN} Dubrovin, B. A. , Krichever, I. M. , Novikov, S. P.
{\em Integrable Systems I} Encyclopedia of Mathematical Sciences,
Dynamical Systems IV (1985), 173---280

\bibitem[DMN]{DMN} Dubrovin, B. A. , Matveev, V. B. , Novikov, S. P.
{\em Nonlinear equations of the Korteweg-de Vries type, finite zone
linear operators, and Abelian manifolds} Russian Math. Surveys,
{\bf 31} (1976), 55---136

\bibitem[Gar]{Gar} Gardner, C. S. {\em Korteweg-de Vries equation and
generalizations IV} J. Math. Phys. {\bf 12} No:8 (1971), 1548---1551

\bibitem[GGKM]{GGKM} Gardner, C. S. , Greene, J. M. , Kruskal, M. D. ,
Miura, R. M. {\em Method for solving the Korteweg-de Vries equation} Phys.
Rev. Lett. {\bf 19} (1967)

\bibitem[G-D]{G-D} Gelfand, I. M. , Dickey, L. A. {\em Fractional powers
of operators and Hamiltonian systems} Func. Anal. Appl. {\bf 10} (1976),
259---273

\bibitem[G-L]{G-L} Gelfand, I. M. , Levitan, B. M. {\em On the
determination of a differential equation from its spectral function} Izv.
Akad. Nauk. SSSR Ser. Mat. {\bf 15} (1951)

\bibitem[Gie1]{Gie1} Gieseker, D. {\em A lattice version of the KP
equation} Acta Math. {\bf 168} (1992)

\bibitem[Gie2]{Gie2} Gieseker, D. {\em The Toda Hierarchy and the KdV
Hierarchy} Commun. Math. Phys. {\bf 181} (1996), 587---603

\bibitem[Gr]{Gr} Green, L. {\em An algebraic geometry approach to
nonlinear difference equations} (preprint)

\bibitem[G-H]{G-H} Griffiths, P. , Harris, J. {\em Principles of
Algebraic Geometry} Wiley Interscience (1978)

\bibitem[Har]{Har} Hartshorne, R. {\em Algebraic Geometry} Graduate Texts
in Mathematics {\bf 52}, Springer-Verlag (1977)

\bibitem[K-M]{K-M} Karasev, M. V. , Maslov, V. P. {\em Nonlinear Poisson
Brackets, Geometry and Quantization} AMS Math. Monographs, vol {\bf 119}
(1993)

\bibitem[K-T]{K-T} Mc Kean, H. P. , Trubowitz, E. {\em The spectrum
of Hill's equation in the presence of infinitely many bands} Comm.
Pure Appl. Math. {\bf 29} (1976), 143---226

\bibitem[Kis]{Kis} Kisisel, A. U. O. {\em On an algebro-geometric
discretization of the KP hierarchy} preprint arxiv:math.MP/0102024

\bibitem[Kri]{Kri} Krichever, I. M. {\em Methods of algebraic geometry in
the theory of nonlinear equations} Russian Math. Surveys {\bf 32:6}
(1977), 185---213

\bibitem[Lax]{Lax} Lax, P. D. {\em Periodic solutions of Korteweg-de Vries
equation} Comm. Pure Appl. Math. {\bf 28} (1975), 141---188

\bibitem[Mag]{Mag} Magri, F. {\em A simple model of the integrable
Hamilton equation} J. of Math. Phys. {\bf 19} Part 1 (1978), 1156---1162

\bibitem[vM1]{vM1} van Moerbeke, P. {\em Integrable Foundations of String
Theory} Integrable Systems, J. L. Verdier Memorial Conference, World
Scientific (1991)

\bibitem[vM2]{vM2} van Moerbeke, P. {\em The spectrum of Jacobi matrices}
Invent. Math. , {\bf 37} (1976), 45---81

\bibitem[vM-M]{vM-M} van Moerbeke, P. , Mumford, D. {\em The spectrum
of difference operators and algebraic curves} Acta Math. {\bf 143} (1979),
93---154

\bibitem[Pal]{Pal} Palais, R. {\em The symmetries of solitons} Bulletin of
the AMS {\bf 34} No:4 (1997)

\bibitem[Wei]{Wei} Weinstein, A. {\em The local structure of Poisson
manifolds} J. of Diff. Geom.  {\bf 18} (1983), 523---557

\bibitem[Tod]{Tod} Toda, M. {\em Theory of Nonlinear Lattices} Springer
Series in Solid State Sciences {\bf 20} (1989)

\bibitem[Z-F]{Z-F} Zakharov, V. E. , Faddeev, L. D. {\em The Korteweg-de
Vries equation is a completely integrable Hamiltonian system} Func. Anal.
Appl. {\bf 5} (1971), 280---287

\end{thebibliography}
\end{document}